\documentclass[12pt]{article}
\usepackage{graphicx, epsf}
\usepackage{amsmath,amssymb,amsfonts}
\usepackage{color}

\newcommand{\be}{\begin{equation}}
\newcommand{\ee}{\end{equation}}
\newcommand{\bea}{\begin{eqnarray}}
\newcommand{\eea}{\end{eqnarray}}
\def\ie{{\it i.e.~}}
\def\del{\partial}

\def\f{{\rm f}}
\def\Nf{N_{\f}}
\def\D{{\rm D}}

\def\XSB{{\chi{\rm SB}}}
\def\XSR{{\chi{\rm SR}}}
\def\Vol{{\rm{Vol}}}
\def\DBI{{\rm{DBI}}}
\def\CS{{\rm{CS}}}
\def\det{{\rm{det}}}
\def\Tr{{\rm{Tr}}}

\def\a{\alpha}

\def\g{\gamma}
\def\d{\delta}
\def\e{\epsilon}

\def\c{\chi}

\begin{document}

\begin{center}\ \\ 
\vspace{60pt}
{\Large {\bf Chiral Condensates in Finite Density\\ Holographic NJL Model from String Worldsheets}}\\ 
\vspace{40pt}

{\large Mohammad Edalati$^1$ and Justin F. V\'azquez-Poritz$^2$}
\vspace{25pt}

{\it $^1$Department of Physics, University of Illinois at Urbana-Champaign,\\ Urbana IL 61801, USA}
\\{\tt edalati@illinois.edu}\\ [4mm]
{\it $^2$Physics Department, New York City College of Technology,\\The City University of New York, Brooklyn NY 11201, USA}

{\it $^2$Physics Department, The Graduate School and University Center, \\The City University of New York, New York, NY 10016, USA}
\\ {\tt JVazquez-Poritz@citytech.cuny.edu}\\ [4mm]
\end{center}

\vspace{50pt}

\centerline{\bf Abstract}
\noindent We calculate the one-point function of certain chiral operators, known as open Wilson lines, in the holographic Nambu-Jona-Lasinio (NJL) model at finite temperature as well as finite chemical potential for quark (baryon) number density. These operators are holographically dual to Euclidean string worldsheets bounded by the flavor branes.  Their one-point functions can serve as order parameters for chiral symmetry breaking in these models. Analyzing the behavior of the dual string worldsheets with respect to temperature and chemical potential (or quark density) enables us to determine how the order parameter depends on these parameters. 

\newpage
\section{Introduction and Summary}
QCD at finite quark density and temperature has a far more interesting phase structure than at just finite temperature; see \cite{rw0011} for a review. For instance, depending on the number of flavors, new phases such as color-superconductivity and color-flavor locked are believed to emerge at moderate to high densities. While the phase diagram is well understood at zero density and finite temperature, little is known at finite density and temperature, especially in the regime of intermediate densities where QCD is strongly coupled. Needless to say, understanding the phase structure of QCD at moderate densities will help us understand the physics of the quark-gluon plasma produced at RHIC, as well as the physical processes within the core of neutron stars. Unfortunately, lattice simulations are not yet applicable for this regime. Our current field-theoretic knowledge in this direction comes from studying Nambu-Jona-Lasinio (NJL) type models with the hope that they shed light on phases of QCD for which the theory is strongly coupled (and gluon degrees of freedom seem to be irrelevant). See \cite{b0402} for a review of NJL models.

To model phenomena arising within strongly-coupled QCD, there is an emerging alternative approach based on gauge-gravity duality \cite{agmoo9905}. Although it is only for a large number of colors $N_c$ and large 't Hooft coupling $\lambda$ that one can do reliable computations using this duality, experience over the past decade indicates that some of these results may be applicable to QCD, which might lie within the same universality class of the theories being studied. One model which shares a number of common strong-coupling features with QCD, such as confinement and chiral symmetry breaking ($\XSB$), is the Sakai-Sugimoto model \cite{ss0412}, also known as holographic QCD. The model consists of $N_c$ ``color'' $\D4$-branes which intersect $N_\f$ ``flavor" $\D8$-branes and $N_\f$ $\overline{\D8}$-branes at two $(3+1)$-dimensional intersections, taken to be in the $x^{0-3}$ directions. The $\D8$- and $\overline{\D8}$-branes are located at the antipodal points of a circle in the $x^4$ direction.  Imposing anti-periodic boundary conditions for fermions around the circle leaves the gauge bosons of the $4-4$ strings massless but gives mass to their fermionic and scalar modes. At the $\D4$-$\overline{\D8}$ and $\D4$-$\D8$ intersections, there are massless Weyl fermions, denoted $\psi_L$ and $\psi_R$, which come from the Ramond-Ramond (RR) sector of the $4$-$\overline 8$ and $4$-$8$ strings and transform as $(\bf{N_c}, \bf{N_\f}, \bf1)$ and $(\bf{N_c},\bf1,\bf{N_\f})$ of ${\rm U}(N_c)\times {\rm U}(N_\f)\times {\rm U}(N_\f)$, respectively.  The ${\rm U}(N_\f) \times {\rm U}(N_\f)$ gauge symmetry of the flavor branes is the chiral symmetry for these Weyl fermions. At large $N_c$ and large effective four-dimensional 't Hooft coupling $\lambda_{\rm eff}$, in the probe approximation ($\Nf\ll N_c$) the $\D8$- and $\overline{\D8}$-branes are smoothly connected into a U-shaped configuration whose interpretation is that the dual gauge theory, which is already confined, is in a $\XSB$ phase. 

Moving the flavor branes away from the antipodal points of the $x^4$-circle leads to a separation in the scales of $\XSB$ and confinement. Furthermore, taking the radius of the circle to be infinite results in a model without confinement \cite{ahjk0604}. This non-compact version of the Sakai-Sugimoto model is dual to a field theory which can be referred to as the holographic NJL model. This model is interesting in its own right since, like the usual NJL models, it enables one to analyze $\XSB$ holographically in a setting where confinement has been turned off. 

These models have been generalized in a number of different ways, which have been found to continue to exhibit similar behavior to QCD and NJL models. For example, temperature can be added to the Sakai-Sugimoto model \cite{asy0604} and its non-compact version \cite{ps0604} by taking the $\D4$-branes away from the extremal limit. At high enough temperatures, the preferred configuration is that of separated parallel $\D8$- and $\overline{\D8}$-branes, which signals chiral symmetry restoration ($\XSR$) in the dual field theory. Including a finite baryon density in the field theory is another generalization of these models which was studied in \cite{ksz0608, ht0608, y0707, bll0708, dgks0708, rsvw0708}. Following \cite{w9805}, the baryons themselves correspond to $\rm{S}^4$-wrapped $\D4$-branes which are dissolved into the flavor branes and appear as their worldvolume instantons \cite{ss0412, hryy0701, hssy0701, ss0810}. The phase diagram of the Sakai-Sugimoto model at finite density and temperature has been  partially mapped out. To some extent, this shows a resemblance to the QCD phase diagram obtained  from analyzing field theoretic toy models.  

Despite a fair amount of success, the Sakai-Sugimoto model and its non-compact version have some serious shortcomings. For instance, one drawback of these models is that one cannot write an explicit mass term for the localized fermions because there is no direction transverse to both the color and flavor branes along which to stretch an open string. Moreover, although these are models of $\XSB$, the order parameter for  spontaneous $\XSB$ is conspicuously absent. A promising proposal for how to modify these models in order to be able to compute the $\XSB$ order parameter as well as incorporate a bare fermion mass has been given by Aharony and Kutasov in \cite{ak0803} (see also \cite{hhly0803} for related ideas). The idea is that, in holographic QCD and the NJL model, the left and right-handed fermions $\psi_{L}$ and $\psi_{R}$ are localized at different points in the $\c$-direction, so that even though $\psi^{\dagger }_{L}\psi_R$ is charged under the chiral symmetry it is not a gauge invariant operator. Hence, $\langle\psi^{\dagger }_{L}\psi_R\rangle$ cannot serve as a $\XSB$ order parameter in these models. However, one can make a gauge-invariant operator out of the left and right-handed fermions by inserting an open Wilson line between them. The vacuum expectation value (vev) of this new operator, hereafter called open Wilson line operator, or OWL operator for short, is an order parameter for $\XSB$. It was argued in \cite{ak0803} that the vev of this operator  can be obtained from the area of a Euclidean string worldsheet bounded by the flavor branes. In other words, this Euclidean worldsheet is the holographic dual of the OWL operator. The vev of the OWL operator was calculated in \cite{ak0803} for the Sakai-Sugimoto and holographic NJL models at zero temperature and density where a non-vanishing result was found. The proposal was then generalized in various directions in \cite{mms0807, aelv0811}, including the effects of sub-leading corrections, temperature and background electric and magnetic fields on the vev of the OWL operator.

Different regions of the phase diagram of QCD are associated with different condensates. To determine, for example, in what regions of the phase diagram chiral symmetry is broken, one has to analyze the behavior of the chiral condensate (the $\XSB$ order parameter) as a function of temperature and density  (or chemical potential) and see whether a non-vanishing condensate minimizes the appropriate thermodynamical potential. As we alluded to above, such analyses cannot easily be carried out for QCD at intermediate densities. Thus, it is interesting to see whether holographic models can provide some results in this direction.

In this paper, we use the proposal of \cite{ak0803} to analyze how the $\XSB$ order parameter in the holographic NJL model responds to turning on quark (number) density, or chemical potential, and temperature. Since this pertains to $\XSB$, we expect the results obtained in this paper to stay qualitatively the same for holographic QCD, as well. This is in analogy to the case in field theory for which NJL models show, at least qualitatively, similar behavior to QCD in regimes where confinement is of minor relevance. 

This paper is organized as follows. In section 2, we review the holographic NJL model at finite temperature and quark density (or equivalently, baryon density, when quark density is divided by $N_c$). We discuss the various sources in the bulk which give rise to a density in the dual field theory.  Along the way, we find a new configuration where $\rm{S}^4$-wrapped $\D4$-branes are at a finite distance from the horizon (in fact, very close to the horizon), and are connected to the flavor branes by finite-length fundamental strings. However, this configuration is unstable against small radial fluctuations in the position of the wrapped $\D4$-branes, and also is less thermodynamically favored compared to other configurations considered  in this section. 

In section 3, we calculate the $\XSB$ order parameter in the presence of finite temperature and density using the proposal of \cite{ak0803}. In the probe approximation, when it comes to calculating the $\XSB$ order parameter, the temperature manifests itself holographically within the background geometry of the nonextremal D4-branes as well as in the boundary condition on the Euclidean worldsheet dual to the OWL operator. The density or chemical potential, on the other hand, only appears through the boundary condition on the worldsheet. We first consider the case of zero temperature and finite quark (or baryon) chemical potential $\mu$. In this case, unlike the usual NJL model, the holographic NJL model cannot realize a $\XSR$ phase, simply because the flavor branes always remain connected. There is, however, a different kind of phase transition at some critical chemical potential $\mu_{\rm cr}$, which is at the order of the baryon mass.  This is a transition from a $\XSB$ phase with zero baryon density, for which the configuration of flavor branes is U-shaped, to a $\XSB$ phase with non-vanishing baryon density whose gravity dual is given in terms of a density of $\rm{S}^4$-wrapped $\D4$-branes attached to the flavor branes \cite{bll0708, rsvw0708}. To leading order in $\lambda_{\rm eff}$, these $\rm{S}^4$-wrapped $\D4$-branes are pointlike \cite{hryy0701, hssy0701}. Working within this approximation we show that, at the phase transition, the $\XSB$ order parameter as a function of chemical potential is continuous, whereas its derivative jumps. Using a combination of analytical and numerical techniques, we show that for chemical potentials above but near $\mu_{\rm cr}$, the $\XSB$ order parameter decreases linearly with $\mu$ but increases for larger $\mu$. For $\mu<\mu_{\rm cr}$, the order parameter is constant and equals its value at zero chemical potential. At finite temperature and density, the situation is more involved. At low enough temperature and chemical potential where the theory is in a $\XSB$ phase with zero density, we find that the $\XSB$ order parameter increases with temperature $T$ as $T^6$, whereas it does not change with $\mu$. Increasing $\mu$ while keeping the temperature low causes the system to hit a critical chemical potential $\mu_{\rm cr}(T)$, beyond which it is in the $\XSB$ phase with nonzero density \cite{bll0708}. Our computations show that in this phase the $\XSB$ order parameter increases with $T$ and initially decreases with $\mu$ but eventually increases with both quantities. At high enough temperatures, the theory is in a $\XSR$ phase with non-zero density. By an explicit calculation involving the global structure of the background geometry, we show that the worldsheet dual to the OWL operator has an infinite area resulting, as expected, in a vanishing $\XSB$ order parameter.   On the $T-\mu$ phase diagram of the holographic NJL model, there is a tri-critical point where three aforementioned phases ($\XSB$ with zero density, $\XSB$ with non-zero density and $\XSR$ with non-zero density) are equally thermodynamically favored. Combining our analyses for each separate phase, we conclude that the $\XSB$ is discontinuous at that point.

\section{Holographic NJL model}

We start this section by briefly reviewing the holographic NJL model and setting the notation that will be used throughout the paper. 

\subsection{Zero temperature and density}
\begin{figure}[h]
 \centerline{\includegraphics[width=4.5in]{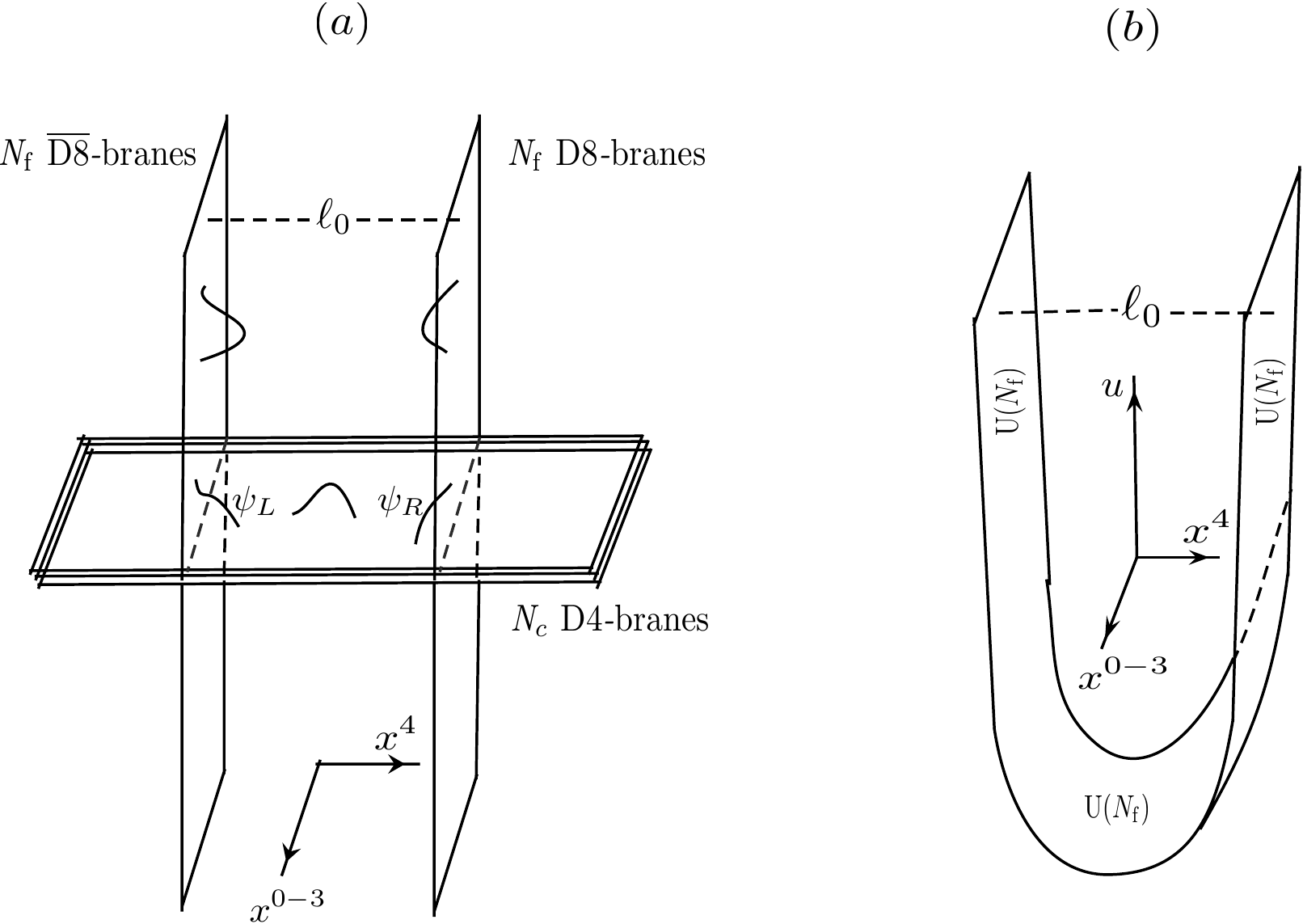}}
   \caption[FIG. \arabic{figure}.]{\footnotesize{(a) The $\D4$-$\D8$ and $\D4$-$\overline \D 8$ intersections in $\mathbb{R}^{1,9}$. (b) At large $N_c$ and large $\lambda_{\rm eff}$, the geometry of color $\D4$-branes is replaced by its near horizon limit and the preferred configuration of the flavor branes is U-shaped, which  signals $\XSB$ in the dual theory.}}
\label{intersection}
\end{figure}

Consider $N_c$ ${\D}4$-branes extended in the $x^{0-4}$ directions in ten-dimensional Minkowski spacetime $\mathbb{R}^{1,9}$. To this system, add $N_{\f}$ $\D8$-branes and $N_{\f}$
$\overline \D 8$-branes such that they intersect the ${\D}4$-branes at two $(3+1)$-dimensional intersections (defects), and are separated in the $x^4$ direction by a coordinate distance $\ell_0$ (see Figure \ref{intersection}a). Focusing on the defects, there are localized massless fermions which come from the RR sector in the $4-\overline 8$ and $4-8$ strings. These fermions are chiral with respect to the ${\rm U}(N_{\f}) \times{\rm U}(N_{\f})$ of the flavor $\overline\D8$- and $\D 8$-branes, and transform as $(\bf {N_{\f}}, \bf {1})$ or $(\bf {1},\bf {{N_{\f}}})$. They are also in the fundamental of ${\rm U}(N_c)$. The left-handed and right-handed fermions, $\psi_L$ and $\psi_R$,  interact via $\D4$-brane gauge fields (and scalars). The strength of the interaction is set by the dimensionless effective coupling  $\lambda_{\rm eff}=\lambda_5/\ell_0$.  

Stringy effects are small when $g_s N_c$ is small. In this regime, when $\lambda_{\rm eff}$ is small one can write down an effective action for the left and right-handed fermions by integrating out the gauge fields of the $\D4$-branes, resulting in \cite{ahjk0604}
\bea\label{intnlnjl}
S_{\rm eff} &=&\int d^4x ~(i\psi^{\dagger}_L\bar\sigma^\mu\del_\mu\psi_L+ i\psi^{\dagger}_R\sigma^\mu\del_\mu\psi_R)\nonumber\\
&+&\frac{g^2_5}{4\pi^2} \int d^4x d^4y G(x-y, \ell_0)\Big[\psi^{\dagger}_L(x)\cdot\psi_R(y)\Big]\Big[\psi^{\dagger}_R(y)\cdot\psi_L(x)\Big], 
\eea
where $G(x-y, \ell_0)$ is a $(4+1)$-dimensional scalar propagator. The dot in the square brackets in (\ref{intnlnjl}) denotes contraction over color indices. The above effective action indicates that  the theory of the fermions at the defects is a NJL model where the four-Fermi interaction term is non-local. In the regime in which the above action is obtained, $\Lambda=\ell_s^{-1}$ is the cut-off of the model. It was argued in  \cite{ahjk0604} that, unlike the local NJL models where spontaneous $\XSB$ happens only above a critical coupling,  chiral symmetry is spontaneously broken at arbitrarily weak coupling for the system of fermions described above.
For large $g_s N_c$ and large $\lambda_{\rm eff}$, the action (\ref{intnlnjl}) can no longer be trusted to provide a reliable description for the dynamics of the fermions.  Instead, this is the regime for which gauge-gravity duality can be used to study the system. In order to do this, one can consider the flavor branes in the near-horizon geometry of $N_c$ extremal $\D4$-branes, described by the metric
\bea\label{Dq metric}
ds^2=\left(\frac{u}{R}\right)^{3/2}\Big(dt^2 + d{\vec x} ^2
\Big) + \left(\frac{u}{R}\right)^{-{3/2}}
\Big(du^2+ u^2 d{\Omega^2_{4}}\Big),
\eea
where $d\Omega^2_4$ is the metric of the unit 4-sphere. The characteristic radius of the geometry $R$ is related to the string coupling $g_s $ and string length $l_s$ by 
\bea\label{R}
R^3=\pi g_sN_c l_s^3=(4\pi)^{-1}g_5^2N_cl_s^2, 
\eea
where $g_5$ is the dimensionful coupling of the low-energy theory of the $\D4$-branes. There is also a dilaton $\phi$ and an RR $4$-form flux  $F_{(4)}$ given by
\bea\label{dilatonRR}
e^{\phi}=g_s\Big(\frac{u}{R}\Big)^{3/4}, \qquad \qquad F_{(4)}=dC_{(3)}=\frac{2\pi N_c}{\Vol({\rm{S}}^4)}\ \e_{(4)},
\eea
where $C_{(3)}$ is a 3-form potential and $\e_{(4)}$ is the volume form of the unit ${\rm{S}}^4$.
The dynamics of the bosonic degrees of freedom of the flavor $\D8$-branes are determined by the Dirac-Born-infeld (DBI) plus the Chern-Cimons (CS) action
\bea\label{dbics}
S=S_{\DBI}+S_{\CS}= -\mu_8\int d^9 \xi~e^{-\phi}  \Tr\sqrt{-\det(g+2\pi\a^{\prime} F)}~+ \mu_8 \int \sum C \wedge \Tr~e^{2\pi\a^{\prime} F}.
\eea
$F_{ab}$  and $g_{ab}$ in (\ref{dbics})  are the field strength and the pullback metric, respectively. The   $\D8$-branes (as well as the $\overline\D8$-branes) are extended in the $\{t, \vec x,  {\rm S}^4\}$ directions and form a curve $u=u(x^4)$ such that the asymptotic distance between the $\D8$- and $\overline\D8$-branes is $\ell_0$: $u(\pm \ell_0/2)=u_\Lambda$. Since the dilaton grows with $u$, $u_\Lambda$ cannot take arbitrarily large values. Indeed, in order for the supergravity approximation to be valid, $u_\Lambda$ has to satisfy $u_{\Lambda} \ll \alpha' N^{1/3} /g_5^2$  \cite{imsy9802}. Since the $\D8$-branes are in the background of $N_c$ $\D4$-branes where only $C_{(3)}$ is turned on, the Chern-Simons part of the action reads 
\be\label{origCS}
S_{\CS}=\frac{\mu_8}{3!}\int C_3 \wedge (2\pi\a^{\prime})^3\ \Tr~F\wedge F\wedge F
=\frac{N_c}{24\pi^2}\int_{{\cal M}^4\times {\rm R}^{+}} \omega_{(5)}(A), 
\ee
where $\omega_{(5)}(A)= \Tr \Big(AF^2-\frac{1}{2}A^3F+\frac{1}{10}A^5\Big)$ is the Chern-Simons $5$-form satisfying $d\omega_{(5)}(A)= \Tr~F\wedge F\wedge F$, and  the integration on the right hand of (\ref{origCS}) is over a five-dimensional space ${\cal M}^4\times {\rm R}^{+}$ spanned by the $x^{0-3}$ and $u$ directions. 

Analyzing the equations of motion coming from (\ref{dbics}), one can set the gauge fields equal to zero. The equation of motion for $u=u(x^4)$ then shows a U-shaped profile for the flavor branes \cite{ahjk0604}. This U-shaped solution, which is energetically favored, is interpreted as representing $\XSB$, where at large $u$  the ${\rm U}(N_{\f})\times {\rm U}(N_{\f})$  symmetry is manifest while at the lowest position of the U-shaped profile  there is just one  ${\rm U}(N_{\f})$ factor (see Figure \ref{intersection}b).  Thus, the holographic NJL model at large $N_c$ and at strong coupling, $\lambda_{\rm eff}\gg1$, has a vacuum in which chiral symmetry is broken.  

\subsection{Finite temperature and density}
At finite temperature, one considers the flavor branes in the near horizon geometry of $N_c$ non-extremal $\D4$-branes whose metric, in Lorentzian signature, is
\bea\label{bhmetricdim}
&&ds^2=\left( \frac{u}{R}\right) ^{\frac{3}{2}}\Big(-f(u)\ dt^2+d{\vec x}^2\Big)+\left( \frac{u}{R}\right)^{-\frac{3}{2}} \Big( \frac{du^2}{f(u)}+u^2 d\Omega_4^2\Big),\\
&&f(u)=1-\frac{u_T^3}{u^3}, \qquad u_T=\Big(\frac{4\pi}{3}\Big)^2R^3T^2,\label{uTdim}
\eea
where $u_T$ is the radius of the event horizon, $T$ is the temperature of the black $\D4$-branes, and  $R$ is given by (\ref{R}). The relationship between $T$ and $u_T$ in (\ref{uTdim})  can easily be understood by Wick rotating the metric in (\ref{bhmetricdim}) to the Euclidean signature and demanding the geometry to be regular at $u_T$. The dilaton and the RR $4$-form flux  are the same as for extremal background, and are given by (\ref{dilatonRR}).

At finite temperature, the analysis of the DBI action for the flavor branes shows two distinct profiles: parallel and U-shaped. The parallel profile represents flavor branes separated from each other going down to the horizon of the background geometry. The field theory interpretation of this solution is that the ${\rm U}(\Nf)\times {\rm U}(\Nf)$ chiral symmetry of the holographic NJL model is unbroken. The U-shaped profile, on the other hand, represents connected flavor branes which stay above the horizon. The dual interpretation of this solution is that  the holographic NJL model is in a phase with spontaneous $\XSB$.  There is a critical temperature of $T_\XSB\simeq 0.15$ (in units of $\ell_0^{-1}$) below which the U-shaped solution is the thermodynamically preferred one\footnote{Strictly speaking, there are two U-shaped solutions at small enough temperatures. One solution bends closer to the horizon while the other one stays farther away from it. It turns out that the solution which stays farther away from the horizon is always thermodynamically favored  \cite{ps0604, bll0708, eln0803}.}. For temperatures above $T_\XSB$, the parallel solution is favored, which results in the restoration of the chiral symmetry of the holographic NJL model. This phase transition is first order  \cite{ps0604}.

The ${\rm U}(1)_{\rm B}$ baryon symmetry in the holographic NJL model is associated with the ${\rm U}(1)_V$ subgroup of the ${\rm U}(\Nf)\times {\rm U}(\Nf)\simeq {\rm SU}(\Nf)\times {\rm SU}(\Nf)\times {\rm U}(1)_V\times{\rm U}(1)_A$ symmetry of the flavor branes. To be more specific, the baryon number is defined to be $N_c^{-1}$ times the charge of ${\rm U}(1)_V\subset {\rm U}(\Nf)_V$. The timelike component of the ${\rm U}(1)$ gauge field on the $\overline\D8$- and $\D8$-branes couples asymptotically to $\psi^{\dagger}_L \psi_L$ and $\psi^{\dagger}_R \psi_R$, respectively. Thus, introducing a chemical potential $\mu$  for the baryon number operator ${\cal Q}_B=N_c^{-1}(\psi^{\dagger}_L \psi_L + \psi^{\dagger}_R \psi_R)$ in the holographic NJL model amounts to turning on the timelike component $A_0(u)$ of the ${\rm U}(1)$ gauge field on the flavor branes and requiring that $A_0(\infty)=\mu$ on both branches. More precisely, according to the standard gauge-gravity dictionary, the non-normalizable mode of $A_0$ gives the chemical potential while the normalizable mode is interpreted as the baryon (quark) number density. In the rest of this paper, unless there is confusion, we will use the baryon number density $n_{\rm B}$ and the quark number density $n_{\rm q}$ interchangeably, and sometimes refer to them simply by the density. It is understood that $n_{\rm B}=N_c^{-1}n_{\rm q}$.

It is convenient to work with dimensionless quantities. We make $\{t, \vec x, u\}$ as well as $A_0(u)$ dimensionless by scaling out factors of $R$ and $2\pi\a^{\prime}$ according to
\bea\label{scale}
\frac{t}{R}\to t, \qquad \frac{\vec x}{R}\to \vec x, \qquad  \frac{u}{R}\to u, \qquad \frac{2\pi\a^{\prime}}{R} A_0(u)\to A_0(u).
\eea
With the above rescaling, the metric takes the form 
\bea\label{bhmetric}
\frac{ds^2}{R^2}=G_{\mu\nu}dx^\mu dx^\nu=u^{\frac{3}{2}}\Big(-f(u)\ dt^2+d{\vec x}^2\Big)+u^{-\frac{3}{2}} \Big(\frac{du^2}{f(u)}+u^2 d\Omega_4^2\Big),
\eea
and $T$ is now the dimensionless temperature given by
\bea\label{uT}
T=\frac{3}{4\pi}\sqrt{u_T}.
\eea
Also, to make the notation less cluttered, we denote the $x^4$ direction by $\c$. Due to symmetry considerations, we choose a purely radial ansatz for the timelike component of the ${\rm U}(1)$ gauge field: $A_0=A_0(u)$.  With flavor branes forming a curve $u(\c)$ in the $\c-u$ plane and with $A_0$ turned on, the action for the total configuration takes the form
\bea\label{flavoractionprime}
S=-2\Nf C\int_{u_t} ^{\infty}du u^{\frac{5}{2}}\sqrt{f(u)u^{3} {\c'}^2+[1-{A_0^{'}}^2] },
\eea
where the primes on $\c$ and $A_0$ denote derivatives with respect to $u$, and $u_t$ is the lowest radius to which the flavor branes descend. Also, the dimensionless constant $C$ is given by
\bea\label{C}
C=\frac{\mu_8}{g_s}R^9\Vol({\rm R}^4)\Vol({\rm S}^4),
\eea
where $\Vol({\rm R}^4)$ is the dimensionless volume of space spanned by the $\{t, x^1, x^2, x^3\}$ directions. Varying (\ref{flavoractionprime}) with respect to $A_0$ and $\c$ results in the following first integrals of motion:
\bea
d&=&\frac{u^{\frac{5}{2}} A_0^{'}}{\sqrt{f(u)u^{3} {\c'}^2+[1-{A_0^{'}}^2]}},\label{d}\\
c_\c&=&\frac{u^{\frac{11}{2}} f(u)\c^{'}}{\sqrt{f(u)u^{3} {\c'}^2+[1-{A_0^{'}}^2]}}\label{cchi},
\eea
where $d$, in some units, is equal to the electric displacement on the the flavor branes and $c_\c$ is a parameter which determines the shape of the flavor branes in the $\c-u$ plane. It is sometimes convenient to trade $A'_0$ in the action (\ref{flavoractionprime}) with $d$ by performing a Legendre transformation on the action. In so doing, we obtain
\bea\label{Legendreaction}
{\cal S}=S-2\Nf C\int_{u_t}^\infty du~A'_0 d=-\Nf C\int_{u_t} ^{\infty}du\ u^{\frac{5}{2}}\sqrt{\Big(1+f(u)u^{3} {\c'}^2\Big) \Big(1+u^{-5}d^2\Big)},
\eea
where we made use of (\ref{d}). Solving for $A_0$ in (\ref{d}) and expanding the result in the asymptotic region $u\to \infty$ yields
\bea
A_0(u)=A_0(\infty) -\frac{2}{3} \frac{d}{u^{3/2}}+\cdots,
\eea
which shows that $d$ appears as the coefficient of the normalizable mode of $A_0$. $d$ is proportional to the baryon number density $n_{\rm B}=\langle{\cal Q}_ B\rangle$ (or, the quark number density $n_{\rm q}$) while  $A_0(\infty)$ equals the dimensionless chemical potential $\mu$ for the density. 

We start the analysis of the above first integrals of motion (\ref{d}) and (\ref{cchi}) by first considering the parallel embedding of the flavor branes for which $\c'=0$, or equivalently $c_\c=0$. In order for the induced metric on the flavor branes not to change sign in this embedding, $A_0$ should satisfy an upper bound of $(A_0^{'})^2<1$. The physical meaning of this bound is straightforward to understand: since $F_{ut}=A_0^{'}$ is the radial electric field on the flavor branes, the bound simply states that $F_{ut}$ must be less than a critical value in order to prevent open strings on the branes from breaking up into pairs. Setting  $\c'=0$ in (\ref{d}), there is a trivial solution $A_0(u)={\rm constant}$ for which $d=0$. Since the near-horizon geometry of $N_c$ non-extremal $\D4$-branes given by the metric (\ref{bhmetric}) is static, the surface $u=u_T$ is a Killing horizon. Indeed, it is a bifurcate Killing horizon at which the timelike Killing vector $\del_t$ vanishes \cite{rw1992}. This, in turn, implies that $A_0$ must vanish at the horizon: $A_0(u_T)=0$. Thus, this configuration of flavor branes represents a chirally-symmetric phase of the holographic NJL model with no density turned on. In order to obtain a nontrivial density, one must turn on a nontrivial (radial) electric field $F_{ut}=A_0^{'}\neq0$ whose lines end on a density of sources.  For a parallel embedding, the flavor branes intersect the horizon, and hence the electric field lines on the branes can end there. Thus, one can think of the horizon as providing a density of sources causing a nontrivial profile for the radial electric field on the branes. Solving (\ref{d}) for $A_0$ for the parallel embedding, we obtain
\bea\label{parallelA}
A_0(u)=\int_{u_T}^u\frac{d}{\sqrt{u^5+d^2}}du, \qquad \Rightarrow \qquad \mu=\int_{u_T}^{\infty}\frac{d}{\sqrt{u^5+d^2}}du,
\eea 
where we set $A_0(u_T)=0$. Note that $A_0$ as given by (\ref{parallelA}) satisfies $(A_0^{'})^2<1$ for $u>u_T$. If we work in the grand canonical ensemble where $T$ and $\mu$ are kept fixed, we then need to compare the grand free energy $\Omega(T,\mu)$ of the two above-mentioned parallel solutions (one with zero density and the other with non-zero density) to see which one is thermodynamically favored. Holographically, $\Omega(T,\mu)$ is given by the Euclidean on-shell action of the configuration of the flavor branes, appropriately renormalized. For parallel embeddings, $\Omega(T,\mu)$ equals the Euclidean continuation of (\ref{flavoractionprime}), once $\chi'=0$ and (\ref{parallelA}) is substituted in (\ref{flavoractionprime}).  Since both of these parallel embeddings have the same UV divergences, one can just subtract their Euclidean on-shell actions, without bothering to renormalize them, and get a finite result. Using (\ref{parallelA}), one obtains that for all $T$ and $\mu$, the parallel configuration with a non-vanishing density is always favored.

For an  embedding where the branes connect, one has $\c' \neq 0$.  There is a solution for which $\c'(u)$, for half of the configuration, say, where $\c \in [0,\ell_0/2]$, is a smooth single-valued function of $u$. In this case, $\c'(u)$ is given by (\ref{cchi}), and $A_0^{'}(u)=0$. This solution represents a $\XSB$ phase of the holographic NJL model with zero density. To be more precise, for $\ell_0$ less than a critical value there are two such solutions, corresponding to two distinct values of $c_\c$. To study the holographic NJL model at finite density, an electric field on the flavor branes must be turned on. To better analyze this case we use the $u=u(\c)$ embedding for which the action for the flavor branes reads 
\bea\label{acnemb}
S=-2\Nf C\int_{0} ^{\ell_0/2}d\c u^{\frac{5}{2}}{\sqrt{f(u)u^{3} +[(\del_{\c}u)^2-{(\del_{\c}A_0)}^2]}}.
\eea
The first integrals of the equations of motion are
\bea
d&=&\frac{u^{\frac{5}{2}} (\del_{\c}A_0)}{\sqrt{f(u)u^{3} +[(\del_{\c}u)^2-{(\del_{\c}A_0)}^2]}},\label{du}\\
c_\c&=&\frac{u^{\frac{11}{2}} f(u)}{\sqrt{f(u)u^{3} +[(\del_{\c}u)^2-{(\del_{\c}A_0)}^2]}}\label{cu}.
\eea
As we alluded to earlier, in order to turn on a baryon chemical potential in the holographic  NJL model, the asymptotic value of $A_0$ must be the same on both branches of the flavor branes \cite{rsvw0708}  which, by virtue of symmetry, means that $A_0$ is symmetric about $\c=0$. Assuming an everywhere smooth $\del_{\c} A_0$, one deduces that $A_0$ at $\c=0$ is an extremum, \ie $\del_{\c} A_0|_{\c=0}=0$. From (\ref{du}), this yields $d=0$ with $A_0$ being a pure gauge everywhere on the branes. In order to avoid this trivial outcome, one can assume that  $\del_{\c} A_0$ is discontinuous at $\c=0$ \cite{dgks0708}, which means that there is a density of charges located at $\c=0$ providing a source (or a sink) for the radial electric field lines. In the presence of this source, a smooth embedding of the flavor branes can no longer be at equilibrium simply because there is no tension in the $u$-direction to counteract the gravitational force exerted by the source. Thus, for a non-vanishing density we discard smooth embeddings as they do not represent stable configurations. A stable configuration may be obtained if the connected embedding has a cusp at some radial point, say at $u=u_t$, in which case $u_t$ and the opening angle of the cusp can be determined by balancing the forces at the cusp. 

\paragraph{Point-like D4-branes} One possible source is fundamental strings attached, at one end, to the flavor branes at $u=u_t$. To obtain a net charge, the string configuration has to connect at the other end to something other than the flavor branes.  One possibility is that the strings end on $\D4$-branes to form a baryon vertex in the background geometry. These $\D4$-branes are extended in the $\{t, {\rm S}^4\}$ directions and are at fixed points in the rest of the directions. The location of the  $\D4$-branes in the $u$ direction can be obtained by extremizing the on-shell action. Equivalently, if we assume that the  $\D4$-branes are at some $u=u_c$, then $u_c$ can be determined by requiring that the total tension in the $u$ direction at $u_c$ vanishes\footnote{This statement assumes that the $\D4$-branes, as well as the strings,  do not interact with one another, and that the strings are all stretched in the $u$ direction.}.  We denote the number of these $\D4$-branes by $2N_{\rm D4}$ and their density by $2n_{\rm D4}$ (where the factors of $2$ are just for later convenience) such that $N_{\rm D4}=n_{\rm D4}\Vol({\rm R}^3)$. Since  the gauge fields on the worldvolume of these $\D4$-branes play no role in our discussion, it is consistent with the equations of motion (obtained from the DBI action) to set them equal to zero. The DBI action of the $\D4$-branes, localized at $u_c$ ($u_T\leq u_c\leq u_t$), then reads 
\bea\label{sourceDactionuc}
S_{\D4}=-\mu_4\int d^5 \xi~e^{-\phi}\  \Tr\sqrt{-\det g_{\D4}}=-2K_{\rm D4}n_{\rm D4}\int du~\d(u-u_c) u \sqrt{f(u)},
\eea
where  $g_{\D4} $ is the induced metric on the $\D4$-branes and $K_{\rm D4}$ is a dimensionless constant given by 
\bea\label{KI}
K_{\rm D4}=\frac{\mu_4}{g_s}R^{5}\Vol({\rm R}^4)\Vol({\rm S}^4).
\eea
There are $N_c$ strings attached to each $\D4$-brane giving a total of $2N_{\rm F}=2N_c N_{\rm D4}$ strings with a density of $2n_{\rm F}= 2N_c n_{\rm D4}$. Again, the factors of $2$ are for convenience in subsequent calculations. The action for $2N_{\rm F}$ strings coupled to $A_0(u)$ on the flavor branes at  $u_t$ and attached  to the $\D4$-branes at $u_c$ is
\bea\label{Fsourceuc}
S_{\rm F}&=&-\frac{N_{\rm F}}{\pi\alpha^{'}}\int d^2 \sigma\sqrt{-\det g_{{\rm F}}} +2N_{\rm F}\int A_\mu dx^\mu\nonumber\\
&=&-2K_{\rm F}N_cn_{\rm D4}(u_t-u_c)+2K_{\rm F} N_c n_{\rm D4}\int A_0(u)\delta(u-u_t)du, 
\eea
where $g_{\rm F} $ is the induced metric on the worldsheet of the strings and $K_{\rm F}$ is a dimensionless constant given by
\bea\label{KF}
K_{\rm F}=\frac{R^2}{2\pi\alpha^{'}}\Vol({\rm R}^4).
\eea
Note that $A_\mu$ and $x^\mu$ in the first line of (\ref{Fsourceuc}) are dimensionful, whereas in the second line their dimensions have been scaled out according to (\ref{scale}).
We denote the tension of the $\D4$-branes in the $u$ direction at $u=u_c$ by $f_{\D4}|_{u_c}$. One can compute it by first varying the on-shell action of the $\D4$-branes with respect to $u_c$ and then multiplying the result by a factor of $[G_{uu}(u_c)]^{-1/2}$ to account for the proper tension \cite{bll0708}. The tension of the strings at $u=u_c$, denoted by $f_{\rm F}|_{u_c}$, can similarly be computed. The magnitude of these tensions takes the form
\bea\label{tend}
f_{\rm D4}|_{u_c}&=&K_{\rm D4}n_{\rm D4} u_c^{\frac{3}{4}}[3-f(u_c)],\\ 
f_{\rm F}|_{u_c}&=&2K_{\rm F}N_c n_{\rm D4}u_c^{\frac{3}{4}}\sqrt{f(u_c)}, \label{tenf}
\eea
where to obtain (\ref{tend}) and (\ref{tenf}) we have assumed that $f(u_c)$ is non-vanishing. Using (\ref{R}) along with $\Vol({\rm S}^4)=\frac{8}{3}\pi^2$,  $\mu_4= 2\pi(2\pi l_s)^{-5}$ and $l_s=\sqrt{\alpha'}$, one deduces  that $3K_{\rm D4}=N_c K_{\rm F}$. 

It is easy to see that, for $u_T<u_c<u_t$,  the two tensions given in (\ref{tend}) and (\ref{tenf}) are equal only at $u_c\simeq 1.08 u_T$. One can show that, when perturbed away from $u_c\simeq 1.08 u_T$, the $\D4$-branes will accelerate further away.  This indicates that the $\D4$-branes are not at a stable equilibrium at $u_c\simeq 1.08 u_T$. A stable configuration can potentially be obtained if the $\D4$-branes are at  $u_c=u_t$. For $u_c=u_t$, the strings have zero tension, whereas the tension of the $\D4$-branes is non-vanishing. Thus, in order to obtain a stable configuration, the non-zero tension of the $\D4$-branes must be equal to the radial tension of the flavor branes at $u_t$. Another potentially stable configuration is obtained when the $\D4$-branes cross through the horizon, so that a bundle of strings stretch between the flavor branes and the horizon. Again, for this configuration to be stable, the tension of the flavor branes at $u_t$ must equal the tension of the strings.

The action for the total configuration, which includes $2\Nf$ flavor branes and a density of  $2n_{\rm D4}$ $\D4$-branes located at $u_t$, reads
\bea\label{totaction}
S&=&-2\Nf C\int_{u_t} ^{\infty}du\ u^{\frac{5}{2}}\sqrt{f(u)u^{3} {\c'}^2+[1-{A_0^{'}}^2]}
-2K_{\rm D4}n_{\rm D4}\int du~\d(u-u_t) u \sqrt{f(u)}\nonumber\\&&+2K_{\rm F}N_c n_{\rm D4}\int A_0(u)\delta(u-u_t)du.
\eea
Setting the variation of the above expression with respect to $A_0$ equal to zero, we obtain
\bea\label{dnd}
d=(\Nf C)^{-1}K_{\rm F}N_c n_{\rm D4}.
\eea
In order for the configuration to be stable, the tension of the $2N_{\rm D4}$ $\D4$-branes at $u_t$ must be equal to the tension  of  the flavor branes. We denote the tension of the flavor branes in the $u$ direction by $f_{\rm flavor}|_{u}$. Due to the symmetry  of the configuration along the $u$-axis, the forces in the $\chi$ direction are already balanced. Keeping $\ell_0$ fixed, the condition for the forces to be equal at $u_t$ is
\bea\label{zfbranes}
2\Nf Cu_t^{-\frac{3}{4}}\sqrt{f(u_t)(u_t^8+u_t^3d^2) -c^2_\chi}=n_{\rm D4} K_{\rm D4}u_t^{\frac{3}{4}}[3-f(u_t)],
\eea
where $c_\chi$ is given by (\ref{cchi}). Note that $n_{\rm D4}$ in (\ref{zfbranes})  is related to $d$ through (\ref{dnd}), and $c_\chi$ is related to $\ell_0$. Thus, for fixed $\ell_0$ (which is our assumption), (\ref{zfbranes}) relates $u_t$ to $d$.

\paragraph{Fundamental strings} As stated before, the configuration of the flavor branes with a density of $2n_{\rm F}=2N_c n_{\rm D4}$ strings stretched from the flavor branes at $u=u_t$ to the horizon is another potentially stable configuration which can source $d$. The action for this configuration is
\bea\label{totactionf}
S&=&-2\Nf C\int_{u_t} ^{\infty}du u^{\frac{5}{2}}\sqrt{f(u)u^{3} {\c'}^2+[1-{A_0^{'}}^2]}-2K_{\rm F}N_cn_{\rm D4}(u_t-u_T)\nonumber\\
&&+2K_{\rm F} N_c n_{\rm D4}\int A_0(u)\delta(u-u_t)du,
\eea
where the orientation of the strings is chosen to be upward, \ie from the horizon to the flavor branes. The relationship between the electric displacement $d$ and $n_{\rm F}$  is the same as the one in equation (\ref{dnd}): $d=(\Nf C)^{-1} K_{\rm F}n_{\rm F}=(\Nf C)^{-1}  K_{\rm F} N_c n_{\rm D4}$.  

Keeping $\ell_0$ fixed, the condition for the forces to be equal at $u_t$ reads 
\bea\label{zfstring}
\Nf Cu_t^{-\frac{3}{4}}\sqrt{f(u_t)(u_t^8+u_t^3d^2) -c^2_\chi}=K_{\rm F}n_{\rm F} u_t^{\frac{3}{4}}\sqrt{f(u_t)}.
\eea
Note that, due to the symmetry of the configuration along $u$-axis, the tensions in the $\chi$-direction are already balanced. Given (\ref{dnd}) and the fact that $c_\chi$ is related to $\ell_0$, (\ref{zfstring}) relates $u_t$ to $d$ for the configuration of the flavor branes and the fundamental strings. 

\paragraph{Instantons (dissolved D4-branes)} Another possible source for $d$ is a density $n_{\rm D4}$ of $\D4$-branes\footnote{Since $\D6$-branes can also end on $\D8$-branes, a bundle of $\D6$-branes stretched between the flavor branes and the horizon is another stable configuration providing a source for $d$. Since a $\D6$-brane ending on $\D8$-branes is a magnetic monopole on the worldvolume of the $\D8$-branes, such a configuration will source a density of magnetic charges in the boundary theory. We do not consider such configurations here because we are mainly interested in turning on sources for objects which are electrically charged under $A_0$, such as quarks.} which are dissolved into the $\D8$-branes and appear as worldvolume  gauge fields carrying instanton charge 
\bea\label{insnumber}
n_{\rm D4}=\frac{1}{8\pi^2}\int_{\cal M} {\rm Tr}F\wedge F.
\eea
The integral in (\ref{insnumber}) is over a four-dimensional space $\cal M$ spanned by $\{x^1, x^2, x^3, u\}$. These instantons couple to the Abelian gauge field on the worldvolume of the $\D8$-branes through the Chern-Simons part of the action.  The electrostatic repulsion among the instantons tends to spread them out and give rise to a distribution of instantons in the $u$-direction, whereas the gravitational force acts in favor of localization and makes the instantons point-like. The competition between these two forces gives the instantons roughly an effective size of $\rho \sim \lambda_{\rm eff}^{-1/2}$ \cite{hryy0701, hssy0701}.  Note that, for $\Nf=1$, finite size instantons cannot be topologically stable and tend to become point-like, whereas for $\Nf>1$ this is not the case. Regardless of the value of $\Nf$, the instantons can effectively be treated as point-like objects for large  enough $\lambda_{\rm eff}$.  Assuming a uniform density $n_{\rm D4}$ of $\D4$-branes localized at $u=u_t$, the Chern-Simons term (\ref{origCS}) reads
\bea\label{CSsource}
S_{\CS}=\frac{R^2}{2\pi\a'}n_{\rm D4}  N_c \Vol({\rm R}^4)\int_{u_t}^{\infty}A_0(u)\d(u-u_t)du,
\eea
which, in turn, results in $d=K_{\rm F} N_c n_{\rm D4}$. Note that $n_{\rm D4}$ is the density of $\D4$-branes needed to source $d$ for half of the flavor configuration. The actual density of instantons for the entire configuration of the $\D8$-$\overline\D8$-branes should therefore be $2n_{\rm D4}$. In the point-like instanton approximation, one can show that the DBI plus Chern-Simons action for the configuration of the flavor branes with a density of (point-like) instantons on their worldvolume is equivalent (up to an $F^2$ term) to the action (\ref{totaction}) for the flavor branes with a density of point-like $\D4$-branes localized at $u_t$ \cite{hryy0701}. We work in this approximation in this paper and can therefore use these two configurations interchangeably. Going beyond this approximation, it turns out that the configuration of flavor branes with a density of finite-size instantons on their worldvolume is thermodynamically favored over the configuration of flavor branes with a density of point-like $\D4$-branes \cite{rsvw0708}.

\section{$\XSB$ order parameter}
\begin{figure}[h]
    \centerline{\includegraphics[width=4.5in]{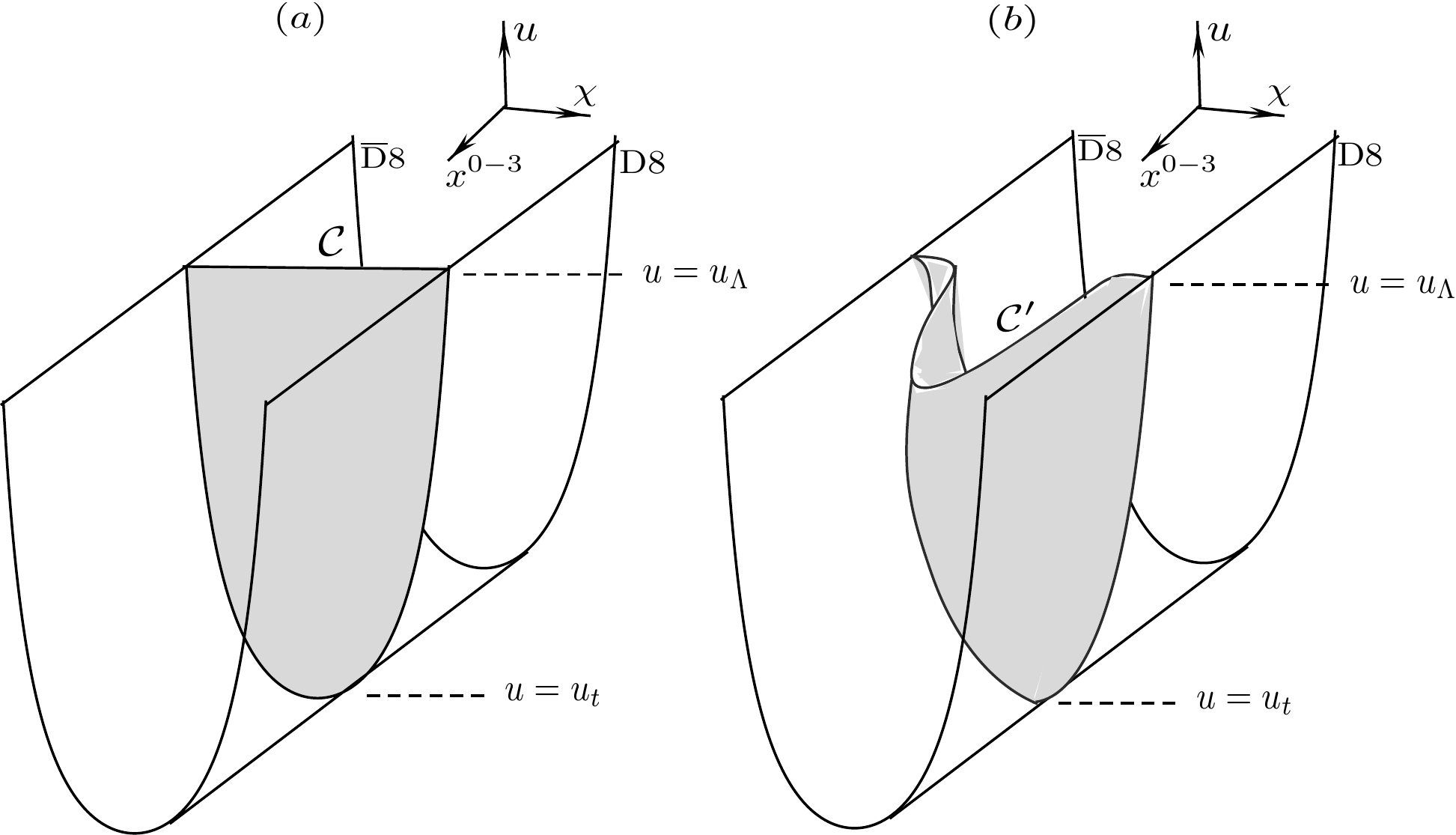}}
   \caption[FIG. \arabic{figure}.]{\footnotesize{(a) The worldsheet dual  to the simplest open Wilson line operator: the one for which the contour $\cal C$ at $u=u_\Lambda$ traverses a straight line in the $\chi$ direction and has a length of $\ell_0$. (b) A worldsheet whose dual open Wilson line operator has a curved contour $\cal C'$.}}
\label{owl-fig}
\end{figure}

In the  holographic NJL model, the left and right-handed fermions $\psi_{L}$ and  $\psi_{R}$ are localized at different points in the $\chi$ direction\footnote{This is also the case for the Sakai-Sugimoto model and, in general, for transversely-intersecting D-brane models considered in \cite{ahk0608, eln0803}.}. At strong coupling, $\lambda_{\rm eff}\gg1$, the model is essentially five-dimensional. Thus, the one-point function of the 
chiral operator $\psi^{\dagger j}_L \psi_{iR}$ cannot be identified as a $\XSB$ order parameter  simply because it is not gauge invariant. Here, $i, j, \cdots$ denote flavor indices. One can make a gauge-invariant chiral operator  by inserting an open Wilson line between 
$\psi_L$ and $\psi_R$. This leads to the so-called open Wilson line operator 
\cite{ak0803}
\bea\label{owl}
{\cal O}^i_j(x^{\mu})=\psi^{\dagger i}_L(x^\mu, \chi=-\ell _0 / 2){\cal P}\ \hbox{exp}\Big[ \int_{-\ell _0 / 2}^{\ell _0 / 2}(i A_\chi + \Phi)d\chi \Big] \psi_{jR}(x^\mu, \chi=\ell _0 / 2),
\eea
where $A_\chi$ is the component of the gauge field in the $\chi$ direction.  The non-local operator ${\cal O}^i_j (x^{\mu})$ depends on the choice of the contour for the open Wilson line insertion. For different contours, one obtains different operators of the above type. The simplest operator is the one for which the contour consists of a straight line traversing the $\chi$ direction. The operator ${\cal O}^i_j (x^{\mu})$ indeed represents a class of non-local chiral operators whose one-point functions can potentially serve as $\XSB$ order parameters of the holographic NJL model (or holographic QCD). One can slightly generalize ${\cal O}^i_j(x^{\mu})$ by considering  $\psi_{L}$ and $\psi_R$ to be located at two different points in the $x^\mu$-directions, in which case we denote the corresponding operator by  ${\cal O}^i_j(x_1^{\mu}, x_2^{\mu})$. In analogy with the well-stablished holographic dictionary for Wilson loops \cite{ry9803, m9803}, it was conjectured \cite{ak0803} that ${\cal O}^i_j(x^{\mu})$, or in general ${\cal O}^i_j(x_1^{\mu}, x_2^{\mu})$, is holographically dual to a Euclidean string worldsheet bounded by the flavor branes as well as the contour of the open Wilson line at $u=u_\Lambda$, while being at fixed points in the rest of the directions\footnote{The authors of \cite{hhly0803} discuss similar ideas on modifying the holographic QCD and NJL models in order to be able to introduce a mass term for the fermions and compute chiral condensates. Refs. \cite{bss0708,dn0708}  (see also \cite{ckp0702}) take another approach for computing fermion masses and chiral condensates in these models, which is based on including the dynamics of open string tachyons in the DBI action of the flavor branes.}. More precisely, the one-point function of the operator is given by
\bea\label{conj-owl}
\langle {\cal O}^i_j \rangle=\delta^i_j e^{-S_{\rm w}},
\eea
where $S_{\rm w}$ is the regularized action of the above-mentioned Euclidean worldsheet. This worldsheet has Neumann boundary condition where it attaches to the flavor branes and Dirichlet boundary condition on the contour at $u=u_\Lambda$. Unless there is  potential for confusion, we will delete the $x^\mu$-dependence of ${\cal O}^i_j(x^{\mu})$ as we already did in writing equation (\ref{conj-owl}). Also, we sometimes omit the flavor indices on the operator and write it simply as ${\cal O}$. Figure \ref{owl-fig}a shows the worldsheet dual to the simplest operator of the type in (\ref{owl}), the one for which the contour $\cal C$ is a straight line extended in the $\chi$ direction at $u=u_\Lambda$. Figure \ref{owl-fig}b depicts the worldsheet dual to an operator with a curved contour $\cal C'$. In practice, it is hard to find  the one-point function of the ${\cal O}^i_j(x^{\mu})$ operators dual to such curved worldsheets. Calculating the one-point function of the simplest open Wilson line operator using the conjecture (\ref{conj-owl}), one then obtains a non-vanishing result  for $\langle{\cal O}^i_j\rangle$ in the vacuum (a $\XSB$ phase) of the holographic NJL model (or holographic QCD) at zero temperature and density \cite{ak0803}.
In this paper, we focus on the aforementioned simplest open Wilson line operator and calculate its one-point function, the $\XSB$ order parameter, in different phases of the holographic NJL model at finite density and finite temperature. We would like to know how  the $\XSB$ order parameter responds to simultaneously turning on chemical potential for the  baryon number density and temperature in the holographic NJL model.  We start our analysis by considering first the case of zero temperature but finite density of the model.

\subsection{Finite density and zero temperature}
At zero temperature but finite density, the configuration of flavor branes and fundamental strings should be discarded. This is because the background geometry does not have an event horizon, in which case the strings have to be stretched from the flavor branes at $u=u_t$ all the way down to $u=0$, where the background curvature (in units of $l_s$) becomes large and the supergravity approximation breaks down. Thus, we are left with either a U-shaped configuration of flavor branes for which  $d=0$, or else a  configuration with finite density for which point-like $\D4$-branes are attached to the flavor branes at $u_t$ and source the finite value of $d$.  On physical grounds, for small enough values of chemical potential, one expects the baryon states not to be able to populate, which results in the solution with zero density being preferred.  Beyond  a critical chemical potential, which is  basically equal to the mass of the baryons (in the approximation where the interactions amongst the baryons are neglected), the expectation is that  the solution with a non-vanishing density is thermodynamically favored. It has been shown in \cite {bll0708, rsvw0708} that such expectations are met in the Sakai-Sugimoto model: for $\mu<\mu_{\rm cr}$, where in the holographic setup $\mu_{\rm cr}$ is the total mass of the point-like $\D4$-branes, the U-shaped flavor configuration has lower free energy, whereas for $\mu>\mu_{\rm cr}$ the configuration of flavor branes with a density of point-like $\D4$-branes localized at $u_t$ is favored. It is easy to show that the holographic NJL model also shows similar behavior. Although the holographic NJL model at zero temperature and finite density behaves qualitatively the same, the field-theoretic NJL models can exhibit rather different behavior.  In particular, at zero temperature, the field-theoretic NJL models undergo a $\XSR$ phase transition at high enough baryon chemical potentials, or densities, (for example, see \cite{b0402}). 

On the other hand, this is not what occurs in the holographic NJL model up to leading order in large $N_c$, since the flavor branes are always connected at zero temperature. Recall that the actions of both point-like $\D4$-branes and flavor branes scale as $N_c$ (to be more precise, $\Nf N_c$), whereas the action for the background geometry scales as $N_c^2$. While the configuration of flavor branes and point-like $\D4$-branes does not back-react on the geometry to leading order up in $N_c$, it does so through the $1/N_c$ corrections. Taking into account the back-reaction, it is plausible that the holographic NJL model may exhibit a $\XSR$ phase at zero temperature but finite chemical potential.  It is interesting to study this particular issue further. In this section, we calculate the $\XSB$ order parameter $\langle {\cal O}^i_j \rangle$ up to leading order in large $N_c$ and large $\lambda_{\rm eff}$ for the above two phases of the holographic NJL model and determine its dependence on both the baryon chemical potential $\mu$ and baryon number density $n_{\rm B}\sim d$.

In order to calculate $\langle {\cal O}^i_j \rangle$, we need to know the profile $u=u(\chi)$ of the flavor branes. Recall that the worldsheet is bounded by the flavor branes defined by the curve $u=u(\chi)$ and by the contour at $u=u_\Lambda$. For $\mu<\mu_{\rm cr}$, the configuration of the flavor branes is U-shaped where the density is zero. Setting $f(u)=1$ and $\partial_\chi A_0=0$ in (\ref{acnemb}), one obtains from the action the first integral of motion
\bea\label{braneeomdzero}
\frac{u^4}{\sqrt{1+u^{-3}(\partial_\chi u)^2}}=u_0^4,
\eea
where $u_0$ is a non-vanishing constant for a U-shaped configuration. Given that $u(\pm \ell_0/2)=u_{\Lambda}$, (\ref{braneeomdzero}) yields
\bea\label{sol}
\ell_0=\frac{1}{4}B \Big( \frac {9}{16}, \frac {1}{2} \Big)\frac{1}{\sqrt{u_0}},
\eea
where to obtain (\ref{sol}) we have ignored terms subleading in $u_{\Lambda}$. We want to calculate the area of a Euclidean worldsheet bounded by the curve $u=u(\chi)$ which is given by (\ref{braneeomdzero}). In the absence of gauge fields on the worldvolume of the flavor branes, the string action is given by
\bea\label{stringaction}
S_{\rm w}=\frac{1}{2\pi \alpha^{'}} \int d^2\sigma \sqrt {\hbox{det} h_{\alpha \beta}}, \qquad 
h_{\alpha\beta}=G_{MN}\del_{\alpha} x^\mu \del_{\beta} x^\nu.
\eea
The string equation of motion then reads
\bea\label{stringeom}
\partial_{\alpha}\partial_{\beta}x^\mu-\Gamma^{\gamma}_{\alpha\beta}\del_\gamma x^\mu+\Gamma^{\mu}_{\nu\rho}\del_\alpha x^\nu\del_\beta x^\rho=0,
\eea
where $\alpha, \beta, \gamma=\{\sigma^1,\sigma^2\}$ denote the worldsheet coordinates while $x^\mu$ are spacetime coordinates.  $\Gamma^{\gamma}_{\alpha\beta}$ and 
$\Gamma^{\mu}_{\nu\rho}$ are the connections that correspond to the induced worldsheet metric $h_{\alpha\beta}$ and the background metric $G_{\mu\nu}$, respectively.
It is straightforward to see that taking
\bea
\sigma^1=R\chi,  \qquad \sigma^2=Ru ,\qquad {\rm and}\qquad x^\mu(\sigma^1,\sigma^2)=x_0^\mu  \qquad (\mu\neq \chi, u),
\eea
where $x_0^\mu$ is a constant, solves the string equation of motion (\ref{stringeom}). This worldsheet, depicted in Figure \ref{owl-fig}a, is the simplest worldsheet and was considered in \cite{ak0803}. The on-shell action of this worldsheet takes the form
\bea\label{ws}
S_{\rm w}&=&\frac{R^2}{2\pi \alpha^{'}} \int du d\chi \sqrt {G_{uu}G_{x^4x^4}}\nonumber\\
&=&\frac{R^2}{2\pi \alpha^{'}} \int d\chi [u_\Lambda-u(\chi)]\nonumber\\
&=&\frac{R^2}{2\pi \alpha^{'}}\ell_0u_\Lambda-\frac{R^2}{8\pi \alpha^{'}}R^{3/2} B\Big( \frac {7}{16},\frac {1}{2} \Big)\sqrt{u_0},
\eea
where $u_\Lambda$ is a cutoff. The linearly diverging piece in the third line of (\ref{ws}) has no effect on our discussion and can, in fact, be absorbed in the definition of $\langle {\cal O}^i_j \rangle$ \cite{ak0803}. This term could either be subtracted away by hand or else done so automatically by performing a Legendre transform on the string action along the lines of \cite{dgo9904} (see, also \cite{cg0812}). Dropping the linear divergent term and using (\ref{R}), (\ref{sol}) and the fact that $\lambda _5= (2\pi)^2 g_s N_c l_s$, we arrive at
\bea\label{zordersw}
S_{\rm w}=-c \lambda_{\rm eff}, 
\eea
where $c=B\Big( \frac {7}{16},\frac {1}{2} \Big) B\Big( \frac {9}{16},\frac {1}{2} \Big)/128\simeq 0.008$. Note that (\ref{zordersw})  is the area of the worldsheet up to leading order in  $\lambda_{\rm eff}$. The subleading  corrections in $\lambda_{\rm eff}$ for this worldsheet have been calculated in \cite{mms0807}. In this paper, we determine $\langle {\cal O}^i_j \rangle$ up to the leading order in $\lambda_{\rm eff}$.  For the $\XSB$ order parameter in a phase of the holographic NJL model where chiral symmetry is broken and the density is zero (or equivalently, for $\mu<\mu_{\rm cr}$), substituting (\ref{zordersw}) into (\ref{conj-owl}) yields
\bea\label{zorderxsb}
{\langle {\cal O}^i_j \rangle}=\delta^i_j e^{c  \lambda_{\rm eff}}.
\eea 
Not surprisingly, the result in (\ref{zorderxsb}) is the same as the one obtained in \cite{ak0803} for the holographic NJL model at zero temperature and zero baryon chemical potential. 

For $d\neq 0$, there is a density of point-like $\D4$-branes on the gravity side attached to the configuration of flavor branes at $u_t$. To find the profile of the flavor branes in this case, we eliminate $\partial_\chi A_0$ between (\ref{du}) and (\ref{cu}) which, having set $f(u)=1$, results in
\bea\label{eqdzerotemp}
(\partial_\chi u)^2=\frac{u^3}{u_*^8}\Big(u^8 +u^3d^2 -u_*^8\Big), 
\eea
where $u_*^8=c_\chi^2$ is a constant. From (\ref{eqdzerotemp}) we obtain
\bea\label{elldzerot}
\frac{\ell_0}{2}=u_*^4\int_{u_t}^{\infty}\frac{du}{u^{3/2}\sqrt{u^8 +u^3d^2 -u_*^8}},
\eea
and from (\ref{zfbranes}) we have
\bea\label{utzerotemp}
u_*^8=u_t^8+\frac{8}{9}u_t^3d^2.
\eea
We are not aware of a closed-form solution for the integral in (\ref{elldzerot}). However, one can perform an analytical approximation in the limit of small $d$. Namely, the integrand in (\ref{elldzerot}) can be expressed as an expansion for $d^2/u_*^5\ll 1$. Defining $v^8=u^8+u^3d^2$, (\ref{utzerotemp}) can be approximated to leading orders to be
\bea\label{approxellzerotemp}
\frac{\ell_0}{2}\simeq \frac{1}{\sqrt{u_*}}\Big[I_3(\epsilon)-\frac{99}{2}\epsilon^2I_{13}(\epsilon)\Big],
\eea 
where
\bea
I_n(\epsilon)=\int_{1+\epsilon^2}^\infty\frac{dv}{v^{n/2}\sqrt{v^8-1}},  \qquad 
{\rm and} \qquad \epsilon^2=\frac{1}{72}\frac{d^2}{u_*^5}.
\eea
Note that, to obtain (\ref{approxellzerotemp}), we have dropped terms subleading in $u_\Lambda$. Expanding the integrals in (\ref{approxellzerotemp}) and keeping terms up to $\epsilon^2$ yields
\bea\label{approxellzerotemp1}
\frac{\ell_0}{2}&\simeq&\frac{1}{\sqrt{u_*}}\Big[I_3(0)-\frac{1}{\sqrt{2}}\epsilon+\frac{99}{2}\epsilon^2I_{13}(0)\Big]\nonumber\\
&=&\frac{1}{\sqrt{u_*}}\Big[\frac{1}{8}B\Big( \frac {9}{16},\frac {1}{2} \Big) -\frac{1}{\sqrt{2}} \epsilon+\frac{99}{16} B\Big( \frac {19}{16},\frac {1}{2} \Big)\epsilon^2\Big]
\nonumber\\
&=&\frac{1}{\sqrt{u_*}}\Big[\frac{1}{8}B\Big( \frac {9}{16},\frac {1}{2} \Big) -\frac{1}{12} u_*^{-5/2}d+\frac{11}{128} B\Big( \frac {19}{16},\frac {1}{2} \Big)u_*^{-5}d^2\Big].
\eea 
For fixed $\ell_0$, (\ref{approxellzerotemp1}) along with (\ref{utzerotemp}) provides a relation between $u_t$ and $d$ for small values of $d$. As is well known, in holographic models the radial direction translates into an energy scale of the dual field theories. In particular, in our holographic setup, $u_t$ can be thought of as the energy scale at which spontaneous $\XSB$ happens. It can be shown from (\ref{approxellzerotemp1}) that this scale decreases with the density $d$ in the limit of small $d$. Numerically, one can verify that this scale keeps decreasing with the density $d$ until $d\approx 0.05$, beyond which it increases. Indeed, this behavior has already been observed in \cite{bll0708}, where a plot of $u_t$ versus density was presented for the Sakai-Sugimoto model at finite temperature and density. 

\begin{figure}[h]
   \centerline{\includegraphics[width=3.5in]{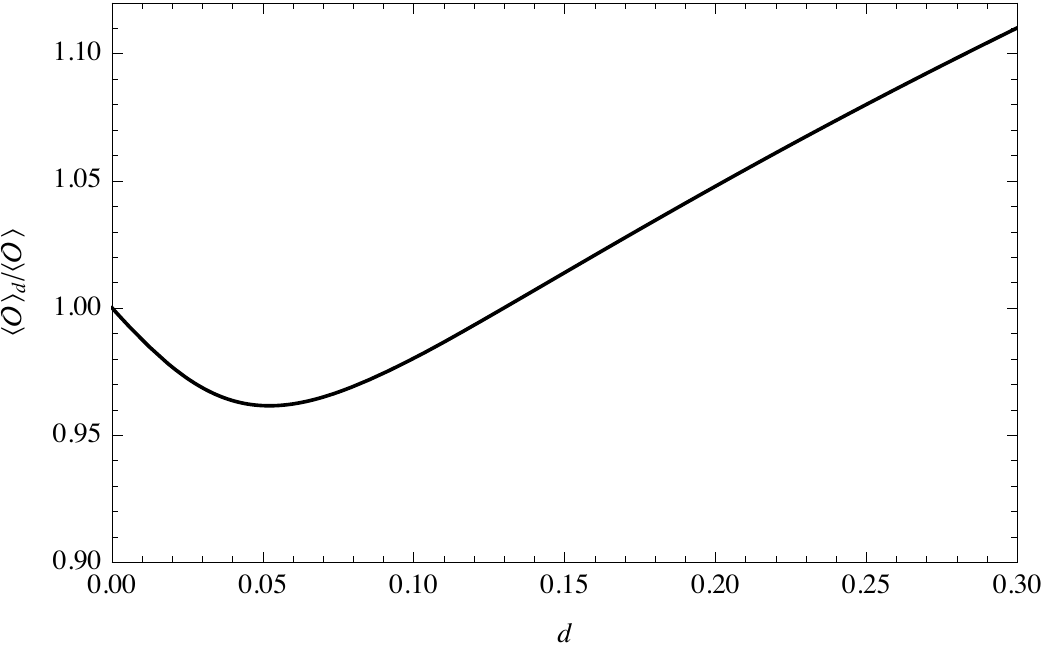}}
   \caption[FIG. \arabic{figure}.]{\footnotesize{$\langle {\cal O}^i_j \rangle_d/\langle {\cal O}^i_j \rangle$  as a function of $d$.  We have set $R=\ell_0=2\pi\alpha'=1$, which all together gives $\lambda_{\rm eff}= 8\pi^2$. We have also chosen the cutoff to be $u_{\Lambda}=10$. The plot is qualitatively the same for other values of $ \ell_0$ and $u_\Lambda$. Although we cut the plot at $d\simeq 0.3$, $\langle {\cal O}^i_j \rangle_d$ is monotonic in $d$ for $d>0.3$.}}
\label{fdowl-one}
\end{figure}

For $d\neq 0$, the string action (\ref{stringaction}) now has an additional boundary term
\bea
S_{\rm b.d.}=\frac{R^2}{2\pi\a'}\int A_0 dt,
\eea
which modifies the string equation of motion (\ref{stringeom}).  But, it can easily be shown that the worldsheet extended in the $\chi-u$ plane and at a fixed point in the rest of the directions is still a solution. Having subtracted the linear divergent term and ignoring terms that are subleading in $u_\Lambda$, the area of this worldsheet reads
\bea\label{areanzd}
S_{\rm w}=-\frac{R^2}{\pi \alpha'} u_*^4\int_{u_t}^\infty\frac{du}{u^{1/2}\sqrt{u^8 +u^3d^2 -u_*^8}}.
\eea
Changing variable to $v^8=u^8+u^3d^2$ and working in the  $d^2/u_*^5\ll 1$ regime, (\ref{areanzd}) is approximated by
\bea\label{approxareazerotemp}
S_{\rm w}&\simeq& -\frac{R^2}{\pi \alpha'}\sqrt{u_*}\Big[I_1(\epsilon)-\frac{81}{2}\epsilon^2I_{11}(\epsilon)\Big]\nonumber\\
&=&-\frac{R^2}{\pi \alpha'}\sqrt{u_*}\Big[\frac{1}{8}B\Big( \frac {7}{16},\frac {1}{2} \Big) -\frac{1}{\sqrt{2}} \epsilon+\frac{81}{16} B\Big( \frac {17}{16},\frac {1}{2} \Big)\epsilon^2\Big]
\nonumber\\
&=&-\Big[\frac{1}{8}B\Big( \frac {7}{16},\frac {1}{2} \Big) -\frac{1}{12} u_*^{-5/2}d+\frac{9}{128} B\Big( \frac {17}{16},\frac {1}{2} \Big)u_*^{-5}d^2\Big].
\eea 
Eliminating $u_*$ between (\ref{approxellzerotemp}) and (\ref{approxareazerotemp}) yields, to leading order,
\bea
S_{\rm w}=-c\lambda_{\rm eff}\Big(1-2.1\ell_0^5 d\Big).
\eea
This results in
\bea\label{owldzerotemp}
\langle {\cal O}^i_j \rangle_d=\langle {\cal O}^i_j \rangle \Big[1-0.02 \lambda_{\rm eff} \ell_0^5 d\Big], 
\eea
where $\langle {\cal O}^i_j \rangle_d$ denotes the $\XSB$ order parameter at finite density. In order to obtain $\langle {\cal O}^i_j \rangle_d$ for arbitrary values of $d$, the integrals in  (\ref{approxellzerotemp}) and (\ref{approxareazerotemp}) should be evaluated numerically. Figure \ref{fdowl-one} shows how $\langle {\cal O}^i_j \rangle_d$ (normalized by its zero density counterpart $\langle {\cal O}^i_j \rangle$) varies with the density $d$. Although not shown here, it could easily be verified that the scale of $\XSB$, which is proportional to $u_t$, exhibits rather similar behavior as a function of the density $d$ as what is shown in the plot of $\langle {\cal O}^i_j \rangle_d$ in Figure \ref{fdowl-one}. Namely, they both decrease with $d$ until $d\approx 0.05$, beyond which they both increase. 
It has previously been found in \cite{aelv0811} that the scale of $\XSB$ and the $\XSB$ order parameter $\langle {\cal O}^i_j \rangle_d$  also share similar behavior in the presence of background electric and magnetic fields in the holographic NJL model:  they both increase (decrease) with a magnetic (electric) field. However, these quantities only have the same qualitative features at zero temperature. In particular, in the holographic NJL model at finite temperature, $u_t-u_T$ decreases with temperature, whereas the $\XSB$ order parameter increases with temperature.

\begin{figure}[h]
      \centerline{\includegraphics[width=3.5in]{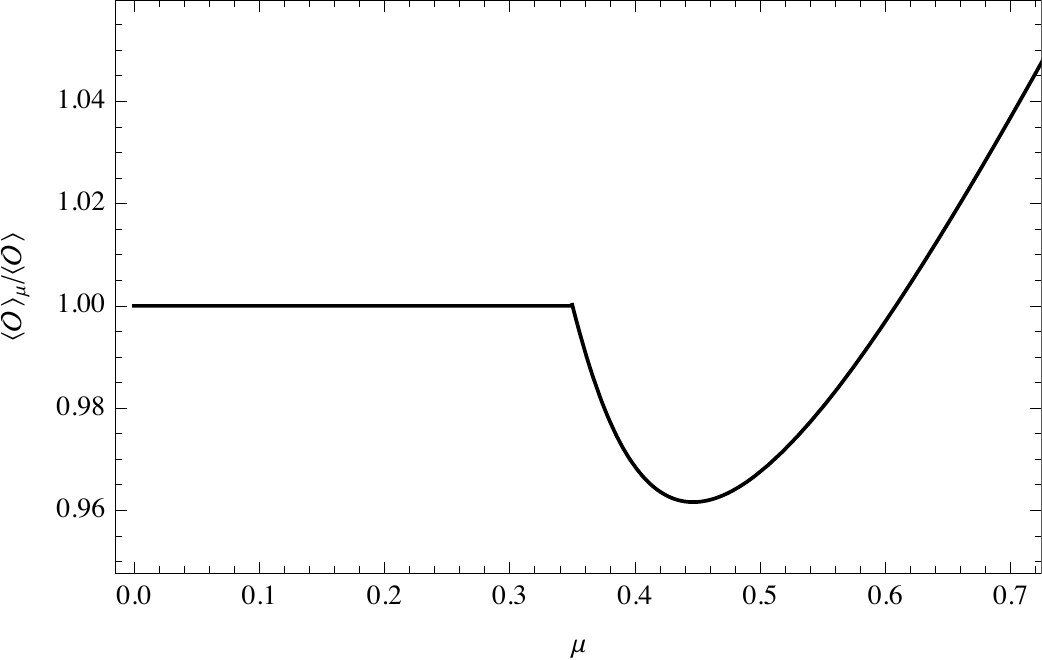}}
   \caption[FIG. \arabic{figure}.]{\footnotesize{$\langle {\cal O}^i_j \rangle_\mu/\langle {\cal O}^i_j \rangle$ as a function of $\mu$. For $\mu<\mu_{\rm cr}\simeq 0.35$, the U-shaped configuration is dominant and $\langle {\cal O}^i_j \rangle_\mu$ does not change with $\mu$. For $\mu>\mu_{\rm cr}$, the configuration of flavor branes with a density of point-like $\D4$-branes is dominant and $\langle {\cal O}^i_j \rangle_\mu$ initially decreases linearly with $\mu$ for $\mu-\mu_{\rm cr}\ll \mu_{\rm cr}$. }}
\label{fdowl-two}
\end{figure}

Given the relationship between the density $d$ and the chemical potential $\mu$, we can use (\ref{owldzerotemp}) to determine the leading-order dependence of the $\XSB$ order parameter on the chemical potential. It was shown in \cite{bll0708,rsvw0708} that, in order to identify $A_0(\infty)$ with the chemical potential, one must identify $A_0(u_t)$ with the total mass of point-like $\D4$-branes localized at $u_t$ (the baryons). Rescaling the chemical potential according to $(2\Nf C) \mu \to \mu$, then from (\ref{d}), (\ref{cchi}) and (\ref{sourceDactionuc}) one obtains 
\bea\label{mudzerotemp}
\mu=\int_{u_t}^\infty 
\frac{u^{3/2}d}{\sqrt{u^8 +u^3d^2 -u_*^8}}du+\frac{R^2}{3\pi\alpha'}u_t,
\eea
where $u_t$ is given in (\ref{utzerotemp}). The value of $\mu$ for $d=0$, denoted by $\mu_{\rm cr}$, is
\bea
\mu_{\rm cr}=\frac{R^2}{3\pi\alpha'}u_*=\frac{1}{192\pi^2}\Big[B\Big( \frac {9}{16},\frac {1}{2} \Big)\Big]^{2}\frac{\lambda_{\rm eff}}{\ell_0}.
\eea
For $\mu<\mu_{\rm cr}$, the U-shaped configuration of the flavor branes is dominant, whereas for $\mu>\mu_{\rm cr}$ the configuration with lower (grand) free energy is made of flavor branes with a density of point-like $\D4$-branes localized on the flavor branes at $u_t$. For $d^2/u_{*}^5\ll1$, the expression in (\ref{mudzerotemp}) approximately equals
\bea\label{muapproxzerotemp}
\mu-\mu_{\rm cr}&\simeq&6\sqrt{2}u_*\epsilon \Big[I_{-3}(\epsilon)-\frac{2\sqrt{2}R^2}{9\pi\a'}\epsilon+\frac{45}{2}I_{7}(\epsilon)\epsilon^2\Big]\nonumber\\
&=&6\sqrt{2}u_*\epsilon\Big[\frac{1}{8}B\Big( \frac {3}{16},\frac {1}{2} \Big) -\Big(\frac{1}{\sqrt{2}}+ \frac{2\sqrt{2}R^2}{9\pi\a'}\Big) \epsilon\Big],
\eea 
where in the second line we have kept terms up to order $\epsilon^2$. Using (\ref{approxellzerotemp}), the expression in (\ref{muapproxzerotemp}) can be rewritten as
\bea\label{mufinalzerptemp}
\mu-\mu_{\rm cr}&\simeq&8 B\Big( \frac {3}{16},\frac {1}{2} \Big) B\Big( \frac {9}{16},\frac {1}{2} \Big)^{-3} \ell_0^3 d.
\eea
Substituting (\ref{mufinalzerptemp}) into (\ref{owldzerotemp}) gives
\bea
\langle {\cal O}^i_j \rangle_\mu=\langle {\cal O}^i_j \rangle \Big[1-0.009 \lambda_{\rm eff}\ell_0^2(\mu-\mu_{\rm cr})\Big],
\eea
for $0<\mu-\mu_{\rm cr}\ll \mu_{\rm cr}$. Note that $\langle {\cal O}^i_j \rangle_\mu$ denotes the $\XSB$ order parameter as a function of chemical potential. Thus, $\langle O^i_j \rangle_\mu$ decreases with $\mu$ for small enough $\mu-\mu_{\rm cr}$ (compared to $\mu_{\rm cr}$). For arbitrary $\mu$, $\langle {\cal O}^i_j \rangle_\mu$ can be calculated numerically. The result is given in Figure \ref{fdowl-two}, where we have again taken $R=\ell_0=2\pi\alpha'=1$ and $u_{\Lambda}=10$. For this choice of parameters, $\mu_{\rm cr}= \frac{1}{24}\Big[B\Big( \frac {9}{16},\frac {1}{2} \Big)\Big]^{2}\simeq 0.35$. Although Figure \ref{fdowl-two} is for particular values of the parameters, we find that the qualitative behavior of $\langle {\cal O}^i_j \rangle_\mu$ is the same for other values of $ \ell_0$ and $u_\Lambda$. Note that at the phase transition which happens at $\mu_{\rm cr}\simeq 0.35$, the $\XSB$ order parameter as a function of chemical potential is continuous, whereas its derivative jumps.

\subsection{Finite density and temperature: $\XSB$ phase}
As we alluded to earlier, at finite temperature but zero density there is a first-order $\XSR$ phase transition at the temperature $T_\XSB\simeq 0.15$ (in units of $\ell_0^{-1}$).  Below $T_\XSB$, the U-shaped solution is energetically favorable and the interpretation is that the chiral symmetry of the holographic NJL model is spontaneously broken. For temperatures greater than the critical temperature, the configuration of parallel branes is thermodynamically favored, which is to say that chiral symmetry gets restored.  It was shown in \cite{aelv0811} that the $\XSB$ order parameter  increases with temperature as the critical temperature is approached from below. In the $\XSR$ phase, on the other hand, the $\XSB$ order parameter vanishes since the dual worldsheet has infinite area. 

\begin{figure}[h]
   \centerline{\includegraphics[width=3.2in]{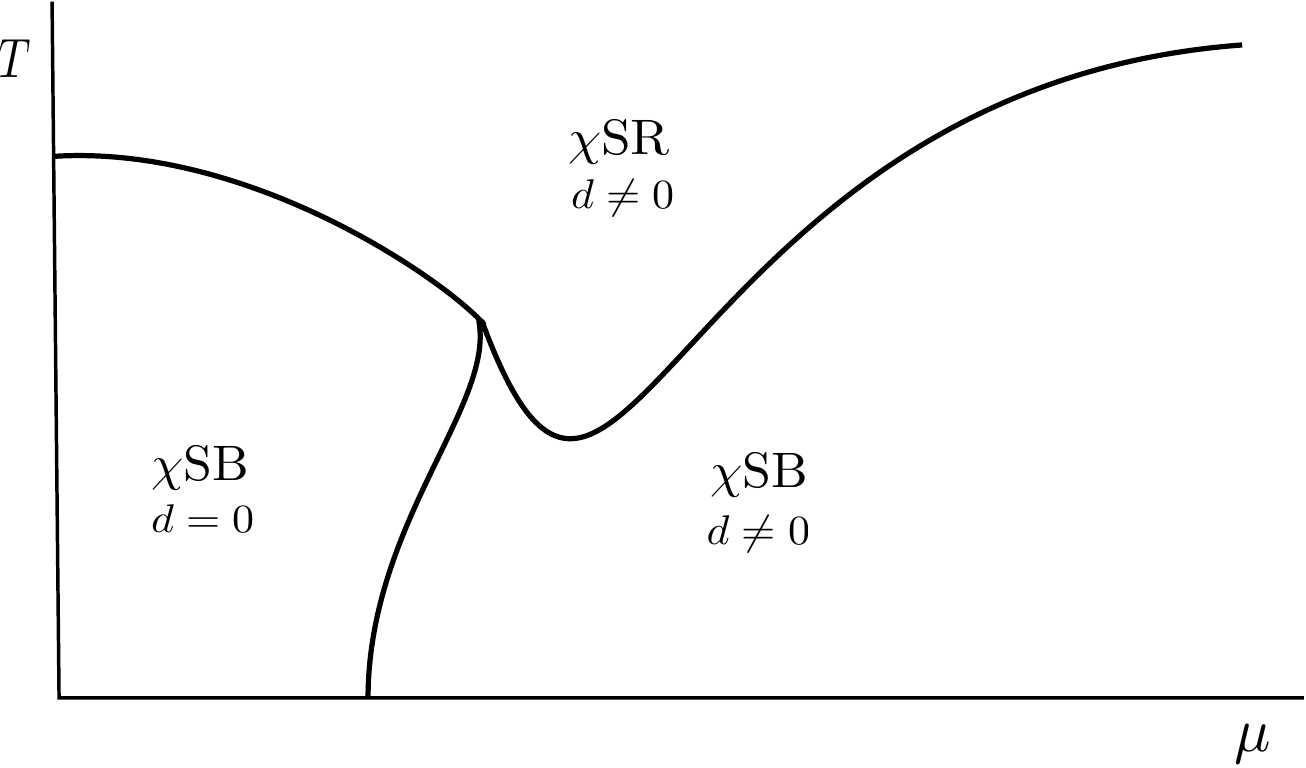}}
   \caption[FIG. \arabic{figure}.]{\footnotesize{A schematic phase diagram of the holographic NJL model at finite temperature and chemical potential.}}
\label{phase}
\end{figure}

At finite temperature and density, the critical temperature above which chiral symmetry is restored decreases with the chemical potential \cite{ht0608, bll0708}. At small enough values of temperature and chemical potential (compared to $\ell_0^{-1}$ and $\mu_{\rm cr}(T)$, respectively), the U-shaped configuration of flavor branes is favored so that the dual field theory is in 
a $\XSB$ phase with no density. Increasing the temperature (while keeping the chemical potential fixed at a low value), one finds that parallel branes supporting a non-trivial density are favored. This implies that the dual field theory is in a $\XSR$ phase at finite density.  Increasing the chemical potential (and keeping the temperature fixed at a low value) results in flavor branes with a density of point-like $\D4$-branes being favored. Thus, the model is in a $\XSB$ phase at finite density. At high enough temperatures and chemical potentials, the configuration of parallel flavor branes which supports a non-trivial density is preferred, implying that the dual field theory is in a $\XSR$ phase at finite density. 
A schematic diagram of the various phases of this model with respect to the temperature and chemical potential is given in Figure \ref{phase}.
A similar diagram has already been obtained numerically in \cite{bll0708} for the Sakai-Sugimoto model. The only difference here is that, for the holographic NJL model, there is no confinement-deconfinement phase transition. 

We will denote the $\XSB$ order parameter at finite density and finite temperature by $\langle {\cal O}^i_j \rangle_{d,T}$. In order to calculate $\langle {\cal O}^i_j \rangle_{d,T}$ at finite temperature and density, we note that the profile of flavor branes satisfies 
\bea\label{braneeomdfinite}
u^4f(u)\sqrt{\frac{1+u^{-5}d^2}{1+u^{-3}(\partial_\chi u)^2}}=u_*^4,
\eea
where $u_*$ is a constant. Given the boundary condition $u(\pm \ell_0/2)=u_{\Lambda}$,  one obtains from (\ref{braneeomdfinite}) (ignoring terms subleading in $u_\Lambda$) that
\bea\label{elldfinite}
\frac{\ell_0}{2}=u_*^4\int_{u_t}^{\infty}\frac{du}{u^{3/2}\sqrt{f(u)[f(u)(u^8 +u^3d^2) -u_*^8]}},
\eea
where $u_t$ (which satisfies (\ref{zfbranes})) is implicitly given by
\bea\label{utfinitetemp}
u_*^8=f(u_t)[u_t^8+u_t^3d^2]-\frac{1}{36}u_t^3d^2[3-f(u_t)]^2.
\eea

It is easy to show that the string equation of motion (\ref{stringeom}) has a solution with a worldsheet extended only in the $\chi-u$ plane and at a fixed point in all other directions. The regularized area of this worldsheet (with the linear divergent term subtracted) reads 
\bea\label{areafinite}
S_{\rm w}=-\frac{R^2 }{\pi \alpha'} u_*^4\int_{u_t}^\infty\frac{du}{u^{1/2}\sqrt{f(u)[f(u)(u^8 +u^3d^2) -u_*^8]}}.
\eea
At low temperatures and small densities, the configuration of flavor branes with a density of point-like $\D4$-branes is dominant. This results in the holographic NJL model being in a $\XSB$ phase at finite density. In this regime,  $d^2/u_{\ast}^5\ll 1$ and $u_T/u_{\ast}\ll 1$. Defining $v^8=f(u)(u^8+u^3 d^2)$ enables one to approximate $\ell_0$ and $S_{\rm w}$ in (\ref{elldfinite}) and (\ref{areafinite}) as
\be\label{approxellwithtemp}
\frac{\ell_0}{2}\approx\frac{1}{\sqrt{u_*}}\left[\frac{1}{8}B\left( \frac {9}{16},\frac {1}{2} \right) -\frac{1}{12} u_*^{-5/2}d+\frac{1}{128}B\left( \frac {15}{16},\frac {1}{2} \right)u_*^{-3}u_T^3\right],
\ee
and
\bea\label{approxareawithtemp}
S_{\rm w}\approx\frac{R^2}{\pi \alpha^{'}}\sqrt{u_*}\left[\frac{1}{8}B\left( \frac {7}{16},\frac {1}{2} \right) -\frac{1}{12} u_*^{-5/2}d-\frac{1}{128}B\left( \frac {13}{16},\frac {1}{2} \right)u_*^{-3}u_T^3\right],
\eea
respectively. Eliminating $u_*$ between (\ref{approxellwithtemp}) and (\ref{approxareawithtemp}) gives
\bea
S_{\rm w}=-c\lambda_{\rm eff}\left[1-2.1\ell_0^5~d+136.2 \ell_0^6~T^6\right],
\eea
where we have used (\ref{uT}) to express $u_T$ in terms of the dimensionless temperature $T$. This results in
\bea\label{owldwithtemp}
\langle O^i_j \rangle_{d,T}=\langle O^i_j \rangle \left[1-0.02 \lambda_{\rm eff} \ell_0^5~d+1.09 \lambda_{\rm eff} \ell_0^6~T^6\right]. 
\eea
Thus, for small values of temperature and density, the $\XSB$ order parameter of the holographic NJL model decreases (linearly) with density and increases with temperature as $T^6$. 

The relationship between $\mu$ (rescaled by $2\Nf C\mu\to \mu$) and $d$ at finite temperature reads
\bea\label{mudwithtemp}
\mu=\int_{u_t}^\infty 
\frac{u^{3/2}d}{\sqrt{f(u)(u^8 +u^3d^2) -u_*^8}}du+\frac{u_t}{3\pi\alpha'},
\eea
where $u_t$ is given in (\ref{utfinitetemp}). The value of $\mu$ at $d=0$ and low temperature is
\bea
\mu_{\rm cr}(T)\simeq\frac{\lambda_{\rm eff}}{192\pi^2\ell_0}\left[B\left( \frac {9}{16},\frac {1}{2} \right)\right]^{2}\left[ 1+512~B\left( \frac {15}{16},\frac {1}{2} \right)B\left( \frac {9}{16},\frac {1}{2} \right)^{-7}\left( \frac{4\pi\ell_0}{3}\right)^6T^6\right],
\eea
from which we see that, at least in the limit of small temperatures, the critical value of $\mu$ increases when the temperature is turned on. For $\mu<\mu_{\rm cr}(T)$, the U-shaped configuration is dominant, whereas for $\mu>\mu_{\rm cr}(T)$ the dominant configuration is made of flavor branes with a density of point-like $\D4$-branes. 
\begin{figure}[h]
    \centerline{\includegraphics[width=3.5in]{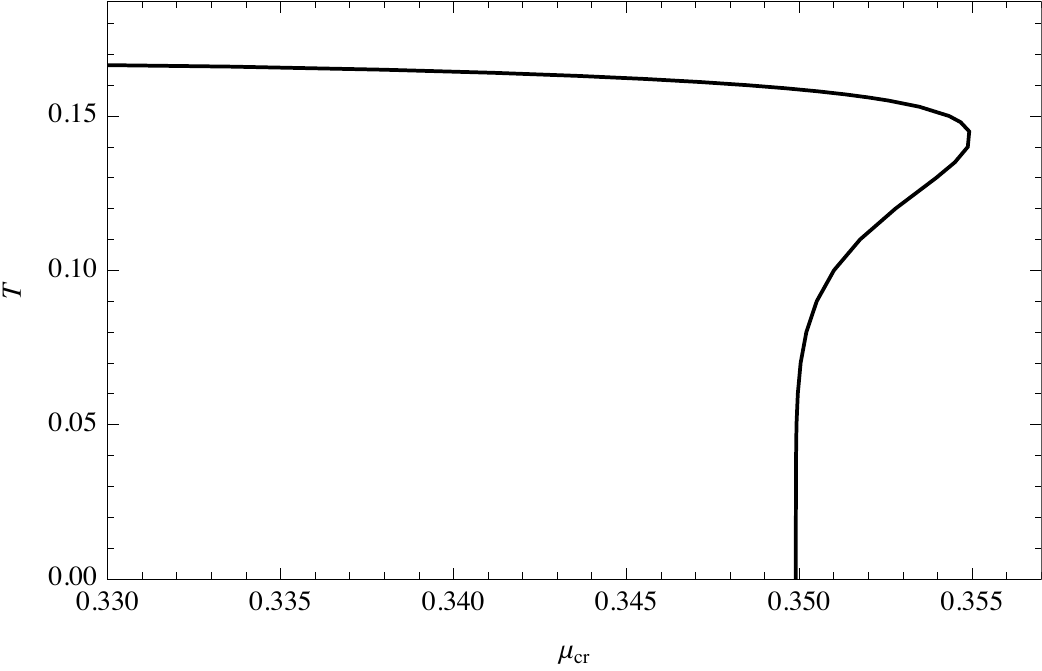}}
   \caption[FIG. \arabic{figure}.]{\footnotesize{Plot of $\mu_{\rm cr}$ as a function of $T$.}}
\label{fdowl-four}
\end{figure}
Figure \ref{fdowl-four} is a numerical result which shows how $\mu_{\rm cr}(T)$ depends on the temperature. Although Figure \ref{fdowl-four} shows $\mu_{\rm cr}(T)$ dramatically decrease for larger temperatures, this does not actually occur since this would only take place for temperatures which are greater than where the $\XSR$ phase transition happens.

\begin{figure}[h]
$\begin{array}{c@{\hspace{0.07in}}c}
\includegraphics[width=3.2in]{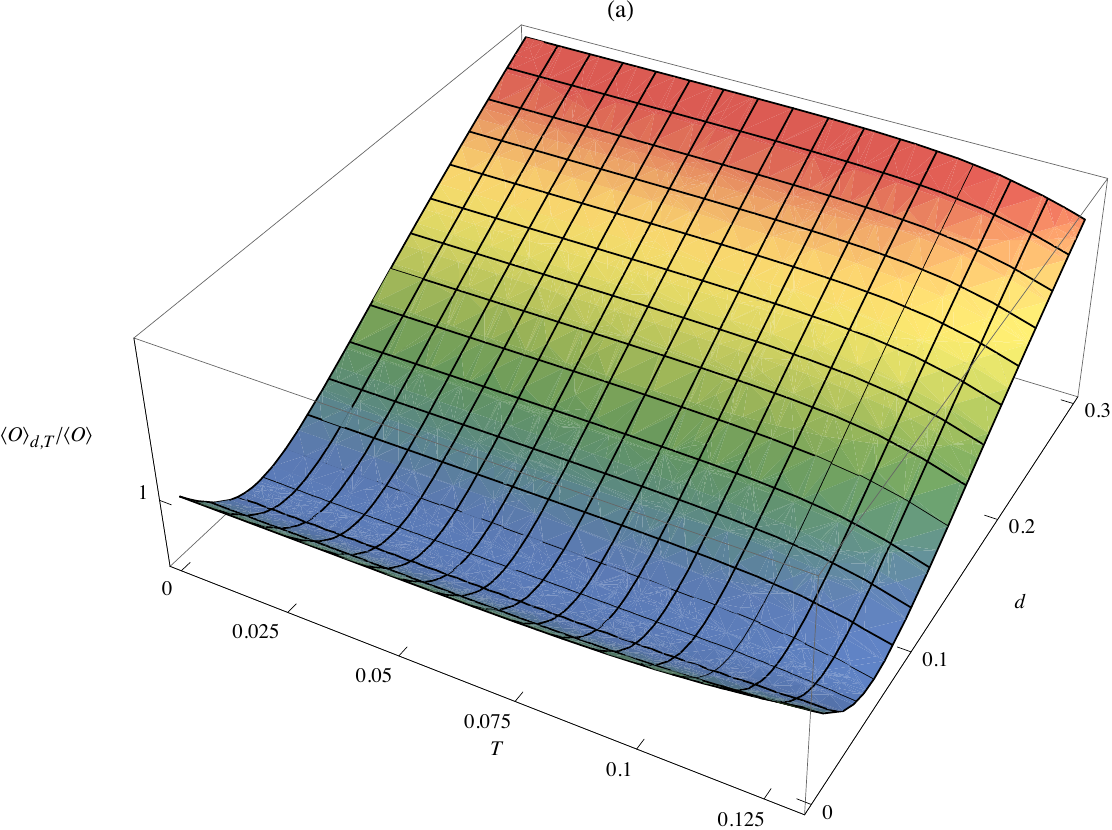}&
\includegraphics[width=3.2in]{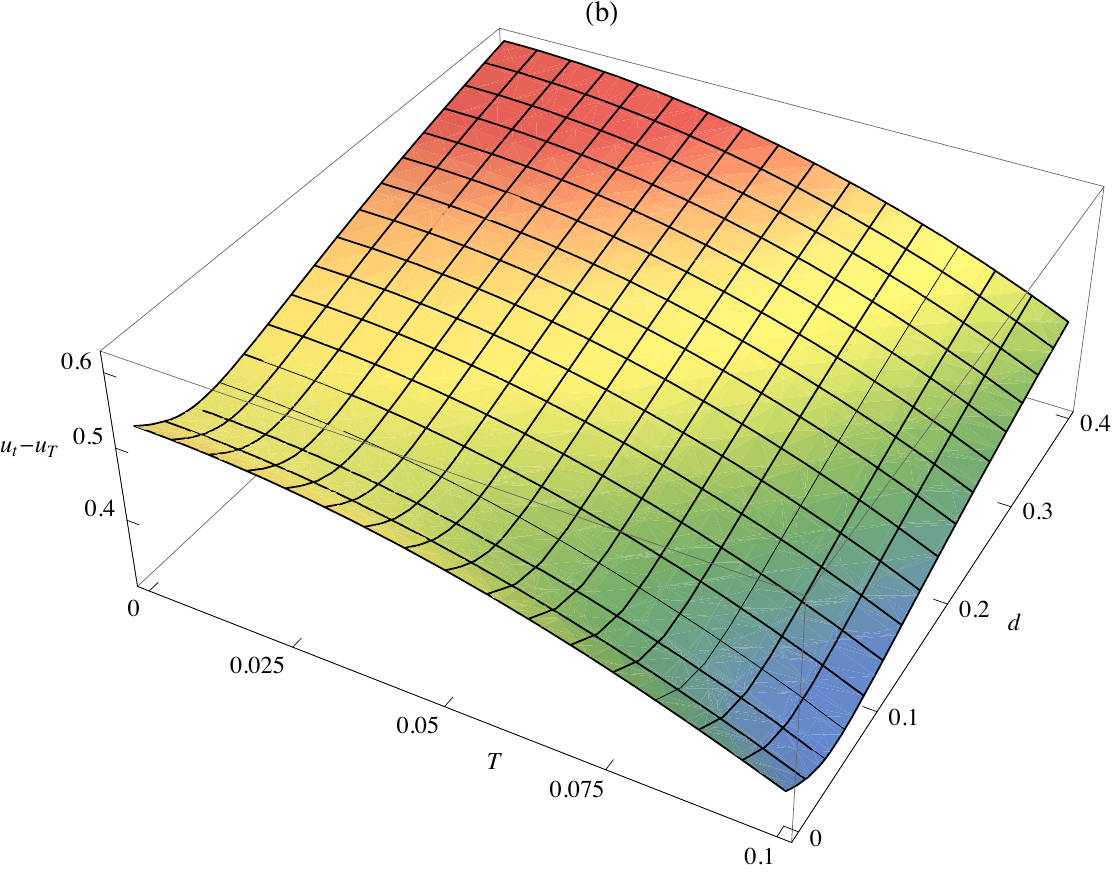}
\end{array}$
\caption[FIG. \arabic{figure}.]{\footnotesize{(a)  $\langle {\cal O}^i_j \rangle_{d,T}/\langle {\cal O}^i_j \rangle$ as a function of $T$ and $d$ in a $\XSB$ phase of the holographic NJL model at low temperatures but arbitrary density. (b)  $u_t-u_T$ as a function of $T$ and $d$ in the same phase.}}
   \label{fdowl-three}
\end{figure}

Since a non-zero temperature does not alter the relation (\ref{mufinalzerptemp}) between $\mu$ and $d$ at the lowest order, we can rewrite (\ref{owldwithtemp}) in terms of $\mu$ as
\bea
\langle {\cal O}^i_j \rangle_{\mu,T}=\langle {\cal O}^i_j \rangle \left[1-0.009 \lambda_{\rm eff}\ell_0^2~(\mu-\mu_{\rm cr})+1.09 \lambda_{\rm eff} \ell_0^6~ T^6\right], 
\eea
for $\mu-\mu_{\rm cr}\ll \mu_{\rm cr}$. Keeping the temperature low while increasing the density, the system stays in the same phase. In order to obtain $\langle O^i_j \rangle_{d,T}$ in this regime,  the integrals in (\ref{elldfinite}) and (\ref{areafinite}) should be evaluated numerically. Figure \ref{fdowl-three} shows the behavior of the $\XSB$ order parameter $\langle {\cal O}^i_j \rangle_{d,T}/\langle {\cal O}^i_j \rangle$ as well as $u_t-u_T$ at low temperatures but arbitrary density. As was the case for the zero temperature, at fixed temperature both of these quantities decrease with $d$ for small $d$ and then increase for larger $d$. On the other hand, at a fixed small density $d$, $\langle {\cal O}^i_j \rangle_{d,T}/\langle {\cal O}^i_j \rangle$ increases with temperature, whereas $u_t-u_T$ decreases with temperature at a fixed density of arbitrary value. However, from Figure \ref{fdowl-three}a it appears that $\langle {\cal O}^i_j \rangle_{d,T}/\langle {\cal O}^i_j \rangle$ actually decreases with temperature for a fixed density at a larger value. Thus, the behavior of $\langle {\cal O}^i_j \rangle_{d,T}/\langle {\cal O}^i_j \rangle$ and $u_t-u_T$ are qualitatively the same (opposite) for smaller (larger) densities.

\subsection{Finite density and temperature: $\XSR$ phase}
At high enough temperatures and densities, parallel branes which support a non-vanishing density are thermodynamically favored. Thus, the holographic NJL model at high temperature and density is in a $\XSR$ phase. Since ${\cal O}^i_j$ is charged under chiral symmetry, one should be able to show that $\langle {\cal O}^i_j \rangle_{d,T}$ vanishes identically for any contour at $u=u_\Lambda$, given that  the string boundary conditions on parallel flavor branes are preserved. Although it is challenging to explicitly show that $\langle {\cal O}^i_j \rangle_{d,T}$ vanishes in general for any contour, in this section we consider a contour that is straight along the $\chi$ direction and analyze the corresponding worldsheets in the $\XSR$ phase of the model. The string equation of motion admits a solution for which the worldsheet only extends in the $u$ and $\chi$ directions and is located at  fixed points in the rest of the directions. Such a worldsheet does not couple to $A_0$ on the flavor branes and is the simplest possible worldsheet one can study.  Since the flavor branes do not form a U-shaped configuration above the horizon, this worldsheet has to go through the horizon. To better analyze the worldsheet, we first consider the situation for which the background geometry (\ref{bhmetric}) has  Minkowski signature. We then consider the worldsheet in the Wick-rotated version of the background geometry. This will enable us to compute the thermal expectation value of the dual operator in the $\XSR$ phase (with finite density) of the holographic NJL model.

In order to study the worldsheet bounded by parallel branes in the background geometry (\ref{bhmetric}), it is convenient to use Kruskal coordinates since this enables us to see what happens to the worldsheet as it passes through the horizon. First, let's scale $u\to u_Tu$ and $x^\mu\to u_T^{-1/2}x^\mu$ in (\ref{bhmetric}) so that the horizon is now located at $u=1$. Next, define the Kruskal coordinates by 
\bea
v = +e^{3 (t+r)/2},\qquad \qquad w = -e^{-3(t-r)/2},
\eea
where $r$ is related to $u$ through 
\be\label{tort}
\frac{dr}{du} = \frac{u^{3/2}}{u^3-1}.
\ee
Note that $r\approx {1\over3} \ln (u-1)$ near $u=1$, so in the new coordinate the horizon is located along the $v$ and $w$ axes. The singularity, on the other hand, corresponds to the curve $vw=1$. In Kruskal coordinates, the background metric takes the form 
\bea
\frac{ds^2}{\gamma^2}=-p(v,w)dvdw+q(v,w)d\vec x^2+ s(v,w)d\Omega^2_4,
\eea
where $\gamma^2=\frac{4\pi}{3}TR^2$, and $p(v,w)$, $q(v,w)$ and $s(v,w)$ are related to the Schwarzchild-like coordinate $u$ through
\bea
p(v,w)=\frac{4}{9}\frac{u^3-1}{u^{3/2}},\qquad q(v,w)=u^{3/2},\qquad s(v,w)=u^{1/2}.
\eea
\begin{figure}[h]
  \centerline{\includegraphics[width=2.5in]{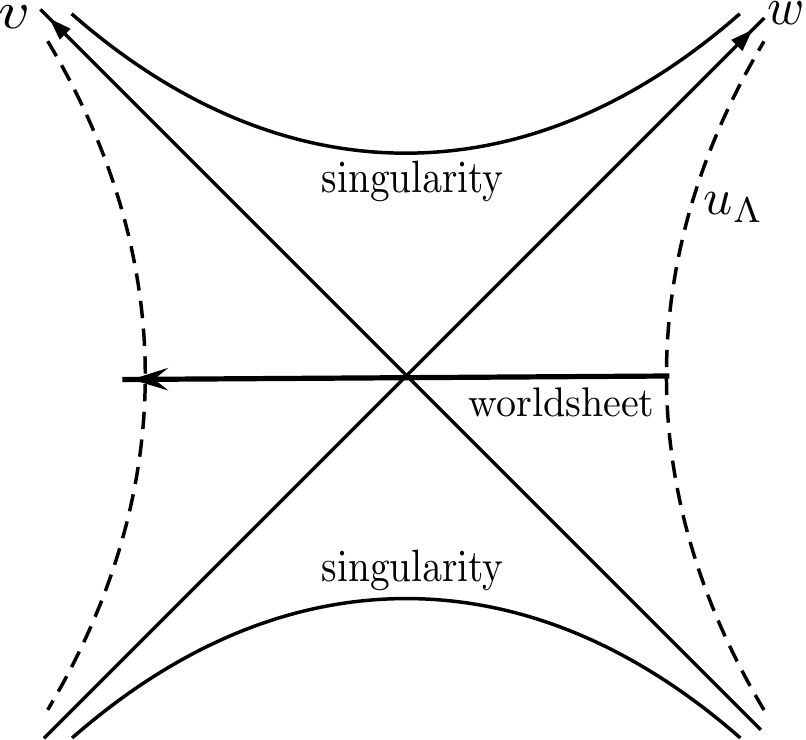}}
   \caption[FIG. \arabic{figure}.]{\footnotesize{The Kruskal diagram of the background geometry. The horizontal line depicts the infinite-area worldsheet. }}
\label{kruskal}
\end{figure}
Since $vw=-e^{3r}$, $p(v,w)$, $q(v,w)$ and $s(v,w)$ are all functions of $vw$ only. As we alluded to earlier, the simplest worldsheet is extended only in the $u$ and $\chi$ directions. Suppose that, with regards to the Schwarzschild time coordinate, the worldsheet is at $t=t_0$. In Kruskal coordinates, this translates into $w = -e^{-3t_0} v$. Since the worldsheet does not depend on time, one can shift $t=t_0$ to $t=0$, which results in the worldsheet being described by the line $w=-v$ in Kruskal coordinates.  Thus, the worldsheet, depicted in   Figure \ref{kruskal} by the horizontal line, passes through the bifurcate horizon and goes to the second asymptotic region in the extended geometry where it has nowhere to end. In other words, the worldsheet has infinite area, which results in $\langle {\cal O}^i_j\rangle=0$.

The analytic continuation of $\langle {\cal O}^i_j\rangle$ to Euclidean signature should be interpreted as a thermal expectation value of the operator. Thus, upon Wick rotation, we should again find that the worldsheet has infinite area, which results in a vanishing thermal expectation value of the dual operator. Let's see this explicitly. To analyze the behavior of the worldsheet, it suffices to just look at the $t-u$ plane, since the $\chi$ direction simply comes along for the ride. Near $u=1$, the Euclidean version ($dt^2\to -dt^2_E$) of the background metric (\ref{bhmetric}) in the $t-u$ plane takes the form 
\bea
\frac{ds^2}{\g^2}=d\rho^2+\rho^2d\theta^2,
\eea
where $\rho=2\sqrt{u-1}/\sqrt3$ and $\theta=3t_E/2$. Since the period of $t_E$ is given by (\ref{uT}), the period of $\theta$ is $2 \pi$ and the geometry is smooth at $\rho=0$ ($u=1$). In the $\{\rho, \theta\}$-coordinates, the worldsheet is located at $\rho_\Lambda$ in the UV region and is at a fixed value of $\theta$, say $\theta=0$. This worldsheet passes through the horizon and emerges at the antipodal point  $\theta=2 \pi/3$, where there is no ${\cal O}$ insertion on which it can end. Equivalently, the worldsheet has infinite area, which results in $\langle {\cal O}^i_j\rangle=0$. This implies that the $\XSB$ order parameter is discontinuous at the tri-critical point on the phase diagram of the holographic NJL model at finite temperature and chemical potential (Figure \ref{phase}), where the three distinct phases of $\XSB$ with zero density, $\XSB$ with non-zero density and $\XSR$ with non-zero density are equally thermodynamically favored.

\section*{Acknowledgments}

We would like to thank S. Baharian, O. Bergman, J.L. Davis, M. Kruczenski, A. Parnachev, M. Rozali and especially P.C. Argyres and R.G. Leigh for illuminating discussions. We are grateful to the Aspen Center for Physics for hospitality during the course of this work. The work of M. E. is supported by DOE grant FG02-91-ER40709.

\end{document}